\documentclass[letterpaper, preprint, paper,11pt]{AAS}	% for preprint proceedings

\usepackage{bm}
\usepackage{amsmath}
\usepackage{amssymb}
\usepackage{gensymb}
\usepackage{threeparttable} 
\usepackage[colorlinks=true, pdfstartview=FitV, linkcolor=black, citecolor= black, urlcolor= black]{hyperref}
\usepackage{overcite}
\usepackage{footnpag} % make footnote symbols restart on each page
\usepackage{wrapfig}
\usepackage{graphicx}
\usepackage{subcaption}
\usepackage{float}
\usepackage{algorithm}
\usepackage{algpseudocode}
\usepackage{enumitem}
\usepackage{booktabs}
\usepackage{makecell}
\usepackage{wrapfig}

\PaperNumber{26-850}

\begin{document}

\title{Stellar Observation Scheduling Optimization for Distributed Space Interferometry Missions}

\author{Ethan Foss\thanks{Ph.D. Candidate, Aeronautics and Astronautics, Stanford University.},  
Antonio Rizza\thanks{PostDoctoral Scholar, Aeronautics and Astronautics, Stanford University.},  
% John Monnier\thanks{Professor, Astronomy, University of Michigan} ,
\ and
Simone D'Amico\thanks{Associate Professor, Aeronautics and Astronautics, Stanford University.},}

\maketitle{}

\begin{abstract}
Growing interest in space interferometry for detecting bio-signatures of exoplanets has led to the development of several low-cost mission concepts involving Earth-orbiting formation flying techniques to perform nulling interferometry of incoming light from exoplanetary systems. Pursuing these developments, this work proposes a multi-objective dynamic programming scheme to optimize the balance between fuel expenditure and scientific outcomes of low-cost formation flying space interferometry missions by leveraging several insights from mission design. This scheme accounts for nominal concept of operations constraints, fuel expenditure, and observability conditions to produce a globally optimal Pareto front policy for exoplanetary system observation, and represents one of the first applications of multi-objective optimization to distributed space telescope astronomy.
\end{abstract}

\section{Introduction}

The search for habitable worlds and extraterrestrial life has led to an explosion of interest in astronomy for detecting and characterizing exoplanets \cite{astro2020_decadal_2023}. Since the first detection of an exoplanet in 1995 \cite{mayor_queloz_1995}, over 6,000  have been detected, with the rate of discoveries steadily increasing \cite{christiansen_et_al_2025}. 

Despite this, there still exist gaps in exoplanet detection and characterization techniques, with most indirect methods of exoplanet detection, such as radial velocity \cite{mayor_queloz_1995}, transit detection \cite{Bryson_2021}, and transmission spectroscopy \cite{Tinetti_Beaulieu_2008} only capable of weakly estimating a limited number of planetary parameters relating to orbital motion, mass, radius, or atmospheric spectrum.

These limitations have motivated the development of direct imaging techniques, which directly image the spectra or radiation signatures of exoplanets\cite{hansen_phd}. The major challenge with these techniques is augmenting the contrast of the target against the brightness of the host star, which has led to the proposal of the techniques of coronagraphy, in which an occulter obscures the incoming starlight from the imaging plane \cite{CRPHYS_2023__24_S2_69_0}, and nulling interferometry, in which incoming starlight is reflected from more than two collectors to a central combiner and destructively interfered to increase the contrast of off-axis objects \cite{hansen_phd}. Though coronagraphy is arguably a more developed concept, with uses in ground telescopes \cite{Jovanovic_2015,beuzit}, space telescopes \cite{Boccaletti}, and even formation flying mission concepts \cite{mdot}, limitations regarding the tradeoff between the inner working angle and contrast place increasing promise on nulling interferometry. In particular, mid-infrared (MIR) nulling interferometry is uniquely poised for the detection of life, with many spectral bio-signatures, such as $O_2$, $H_2O$, and $CO_2$, being within the MIR band \cite{bracewell_1978}. 

The capabilities of MIR nulling interferometry led to the proposal of NASA's Terrestrial Planet Finder (TPF) I and ESA's Darwin missions in the early 2000s, which consisted of 4-5 formation flying spacecraft realizing 10-100 meter separation to measure exoplanetary atmospheres. However, due to low technology readiness levels, these missions were not pursued. Recently, a renewed push to realize MIR space interferometry, following from the proposal of a new mission, LIFE \cite{Quanz_2022}, has led to several developments in the field that warrant further exploration. In anticipation of this mission, several technology demonstrator missions, including SEIRIOS \cite{Ikari2021SEIRIOS}, SILVIA \cite{itosilvia}, and STARI \cite{Monnier2025STARI,Rizza2026}, are in development to realize a low-Earth orbit concept of space interferometry, which was originally theorized by Jonah Hansen in Ref.~\citenum{Hansen_2020}. 

To realize the advanced science and technology needed to perform MIR space interferometry, this paper proposes a scheme that builds off of popular mission concepts \cite{Rizza2026,Hansen_2020,Ikari2021SEIRIOS} to optimize the tradeoff between fuel expenditure and scientific acquisition of low Earth orbit (LEO) linear formation missions. This scheme leverages several concepts to develop a realistic scheme for scheduling starlight measurements, such as the assignment of scientific weights to specific extrasolar systems, simplifications produced by relative orbital elements, and dynamic programming to obtain global Pareto optimality. 

% Though several works have leveraged similar principles of dynamic programming and pareto fronts for tradeoff schemes between scientific acquisition and fuel expenditure, such as Ref.~\citenum{Takubo2024Automated}, which applied dynamic programming to Saturnian moon tour design, and Ref.~\citenum{li2024}, which applied dynamic programming to Earth observation, similar schemes have not yet been applied to optimizing scientific acquisition for space interferometry. Though a scheduling optimization scheme for LEO space interferometry has been proposed in Ref.~\citenum{Hansen_2020} using simulated annealing, this scheme lacks several practical considerations from concept of operations, does not account for fuel optimality, and is suboptimal. 

The proposed scheme, which leverages multi-objective dynamic programming \cite{Bellman1954Theory}, has been applied to several concepts in astrodynamics, including planetary moon tour design \cite{Takubo2024Automated,bellome2023,Brinckerhoff2009,landau2019} and Earth observation \cite{li2024}, though, to the authors' knowledge, has never been applied to observation scheduling of distributed telescopes. Scheduling optimization is commonly studied for astronomy missions \cite{spohnUnderstandingHWOsField2026}; however the unique features of LEO space interferometry missions require tradeoffs and practical considerations, such as multi-faceted constraints and dependency of observation direction on fuel expenditure, that are not adequately captured by existing schemes. Though a scheduling optimization scheme for LEO space interferometry has been proposed in Ref.~\citenum{Hansen_2020} using simulated annealing, this scheme lacks several practical considerations from concept of operations, does not account for fuel expenditure, and is suboptimal. 

The paper is organized as follows: first, background on relevant concepts like relative orbital elements and low-Earth orbit space interferometry, are introduced. Then, constraints on observation are introduced, which inform the set of stars that can be observed and for what duration. Next, the scheduling optimization scheme and its variants are proposed. Finally, the results are presented and discussed.

\section{Background}

In this section, the background on LEO formation flight concepts and their realization in the context of space interferometry missions are introduced. This includes a definition of the relative orbital elements, absolute and relative orbit definitions, and a nominal concept of operations for such a mission. The ensuing analysis considers absolute orbits of the perturbed two-body problem, namely Sun-synchronous orbits, which are of particular interest to mission designs of MIR interferometry.

\subsection{Absolute and Relative Orbital Elements}

Orbital motion of formation flying spacecraft is often described under the definition of relative orbital elements, which provide several dynamical and geometrical insights. In the context of space interferometry, Ref.~\citenum{Rizza2026} has shown that they produce several simplifications for orbital design. Formally, given a chief spacecraft whose orbital elements are given by $ \textit{\textbf{\oe}}_c := \begin{bmatrix}
    a_c, e_c, i_c, \Omega_c, \omega_c, M_c
\end{bmatrix}^\top$ and one or more deputies whose orbital elements are given by $\textit{\textbf{\oe}}_d := \begin{bmatrix}
    a_d, e_d, i_d, \Omega_d, \omega_d, M_d
\end{bmatrix}^\top$, the quasi non-singular relative orbital elements between a deputy and the chief spacecraft are defined as \cite{damico}
\begin{equation}
    \delta \textit{\textbf{\oe}} := \begin{bmatrix}
        \delta a \\ \delta \lambda \\ \delta e_x \\ \delta e_y \\ \delta i_x \\ \delta i_y
    \end{bmatrix} = 
    \begin{bmatrix}
        (a_d-a_c)/a_c \\
        (M_d + \omega_d) - (M_c + \omega_c) +(\Omega_d - \Omega_c) \cos i_c \\
        e_d \cos\omega_d - e_c \cos\omega_c \\
        e_d \sin\omega_d - e_c \sin\omega_c \\
        i_d-i_c \\
        (\Omega_d - \Omega_c) \sin i_c
    \end{bmatrix}.
\end{equation}
\noindent Under this state representation, the linearized equations of motion in the absence of perturbations and under the assumption of affine control $\bm{u}$, can be written as \mbox{$\delta \dot{\textit{\textbf{\oe}}} = A \delta \textit{\textbf{\oe}} + \frac{1}{na}B(t) \bm{u}(t)$}, where 
\begin{equation}
\begin{gathered}
    A = \begin{bmatrix}
        0 & \bm{0}_{1 \times 5} \\
        -\frac{3}{2}n & \bm{0}_{1 \times 5} \\
        \bm{0}_{4\times1} & \bm{0}_{4 \times 5}
    \end{bmatrix}, \
    B(t) =  \begin{bmatrix}
        0 & 2 & 0 \\
        -2 & 0 & 0 \\
        \sin u & 2\cos u & 0 \\
        -\cos u & 2\sin u & 0 \\
        0 & 0 & \cos u \\
        0 & 0 & \sin u
    \end{bmatrix}.
    \end{gathered}
\end{equation}
\noindent Here, $u:= \omega + M$ is the mean argument of latitude of the chief, which satisfies $\dot{u} = n:= \sqrt{\mu/a^3}$. Finally, relative orbital elements can be transformed to relative positions and velocities in the radial-transverse-normal chief-fixed frame with the approximate linear transformation $\delta \bm{x}^\top(t) := \begin{bmatrix} \delta \bm{r}^\top(t) & \delta \bm{v}^\top(t) \end{bmatrix} = a R(t) \delta\textit{\textbf{\oe}}(t)$, where
\begin{equation}
    R(t) = \begin{bmatrix}
           1 & 0 &  -\cos u &   -\sin u &      0 &       0 \\
   0 & 1 & 2\sin u & -2\cos u &      0 &       0 \\
  0 & 0 &        0 &         0 & \sin u & -\cos u \\
  0 & 0 &   \sin u &   -\cos u &     0 &       0 \\
 -\frac{3n}{2} & 0 & 2\cos u &  2\sin u &      0 &       0 \\
   0 & 0 &        0 &        0 & \cos u &  \sin u
    \end{bmatrix}.
\end{equation}

\subsection{Orbit Design}
Mid-infrared space interferometry involves the precise formation flight of three or more spacecraft, of which one is a central ``combiner" that receives starlight that is reflected off of the remaining ``collector" spacecraft. 
%An example of such a concept is depicted in \ref{fig:InterferometryConcept}. 
Of tantamount importance to obtaining a ``deep null," the gold standard of MIR interferometry \cite{hansen_phd}, is precise control of the optical path difference (OPD). OPD refers to the difference in distance that incoming light travels when two or more beams are interfered on the central combiner spacecraft. Obtaining a deep null requires sub-nanometer level OPD \cite{Hansen_2020}, which requires that any formation flying spacecraft achieve centimeter level positioning control (with delay lines). Formally, given an inertial star target, which can be expressed by a unit vector ($\hat{s}_{ECI}$) in the ECI frame or equivalently as right ascension ($\alpha$) and declination ($\delta$)  angles in the equatorial J2000 frame, $\hat{s}^\top_{ECI} = \begin{bmatrix} \cos\alpha \cos\delta & \sin\alpha \cos\delta & \sin\delta \end{bmatrix}$, the OPD ($\Lambda$) of a one combiner, two collector system is \cite{Hansen_2020}
\begin{equation}
    \Lambda = (\| \delta \bm{r}_{ECI,1} \|_2 - \delta \bm{r}_{ECI,1} \cdot \hat{s}_{ECI}) - (\| \delta \bm{r}_{ECI,2} \|_2 - \delta \bm{r}_{ECI,2} \cdot \hat{s}_{ECI}),
    \label{eq:OPD}
\end{equation}
\noindent where $\delta \bm{r}_{ECI,1}$ and $\delta \bm{r}_{ECI,2}$ are the relative positions of the collectors with respect to the combiner. Ideally, a space interferometer should achieve a close to zero OPD continuously during integration and scientific acquisition, which may last for several hours.  

As was discovered in Ref.~\citenum{Hansen_2020} and formalized under a relative orbital element definition in Ref.~\citenum{Rizza2026}, there exists only one class of relative orbits that leverages natural Keplerian relative motion to achieve $\Lambda = 0$ continuously for an arbitrary observation direction under unperturbed linear dynamics. Three-spacecraft formations in these relative orbits have been termed ``linear formations" and can be parameterized in terms of relative orbital elements with
\begin{equation}
    \delta \textit{\textbf{\oe}} = \pm \frac{1}{a}\begin{bmatrix}
        0 & B_0 & 0 & 0 & B_0 \frac{\cos \phi_0}{\tan \beta} & B_0 \frac{\sin \phi_0}{\tan \beta}
    \end{bmatrix}^\top
    \label{eq:scienceroe}
\end{equation}
\noindent or equivalently in RTN positions as
\begin{equation}
    \delta \bm{r} = \pm \begin{bmatrix}
        0 & B_0 & B_0 \frac{\sin(u-\phi_0)}{\tan\beta}
    \end{bmatrix}^\top.
\end{equation}
\noindent Here, $\beta$ and $\phi_0$ are fixed angles that describe the motion of an inertial target in the RTN frame, i.e. $\hat{\bm{s}}^\top_{RTN} = \begin{bmatrix} \cos \beta \cos(\phi_0-u) & \cos\beta \sin(\phi_0-u) & \sin \beta \end{bmatrix}$, and $B_0 >0$ is the minimum baseline distance, for which any value maintains zero OPD. It is easy to verify that $ \hat{\bm{s}}_{RTN} \cdot \delta \bm{r}(t) = 0 \ \forall t$, and therefore $\Lambda(t) = 0 \ \forall t$ under this configuration \cite{Rizza2026,Hansen_2020}.

\begin{figure}[htbp]
    \centering

    % ---------- Row 1 ----------
    \begin{subfigure}{.23\textwidth}
        \centering
        \includegraphics[width=\linewidth]{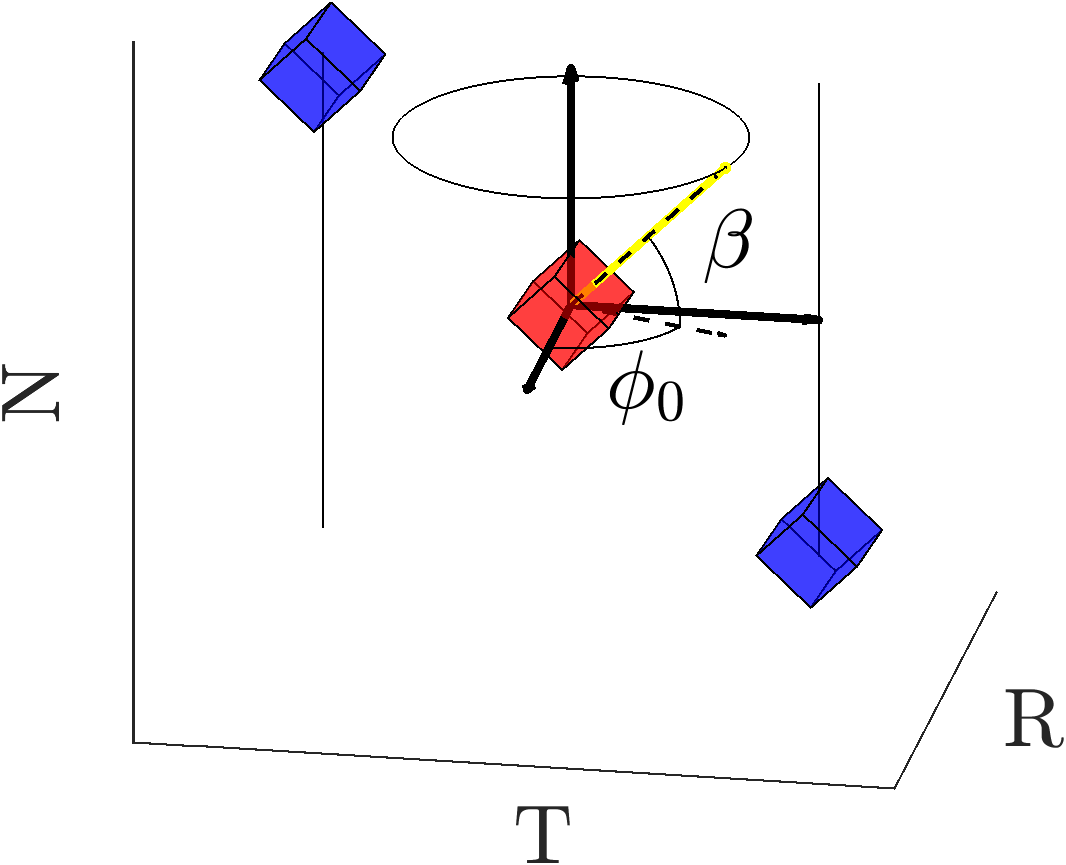}
        % \caption{0 Orbits}
    \end{subfigure}\hfill
    \begin{subfigure}{.23\textwidth}
        \centering
        \includegraphics[width=\linewidth]{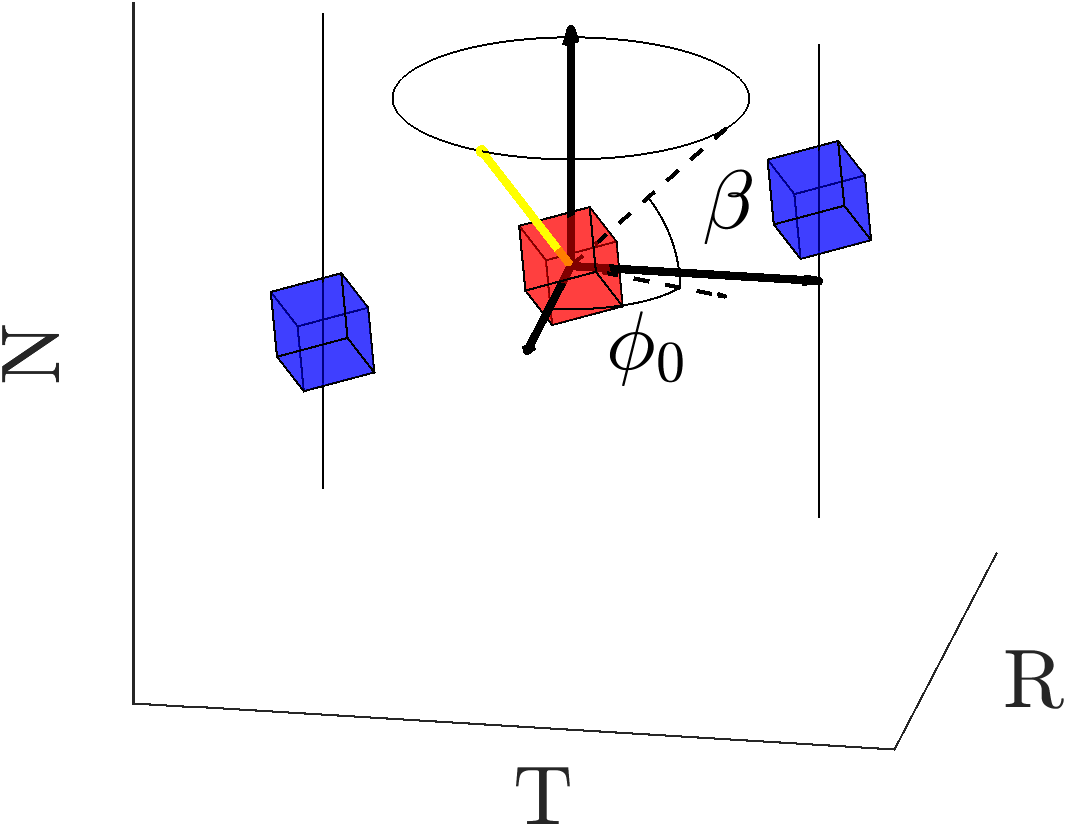}
        % \caption{1/4 Orbits}
    \end{subfigure}\hfill
    \begin{subfigure}{.23\textwidth}
        \centering
        \includegraphics[width=\linewidth]{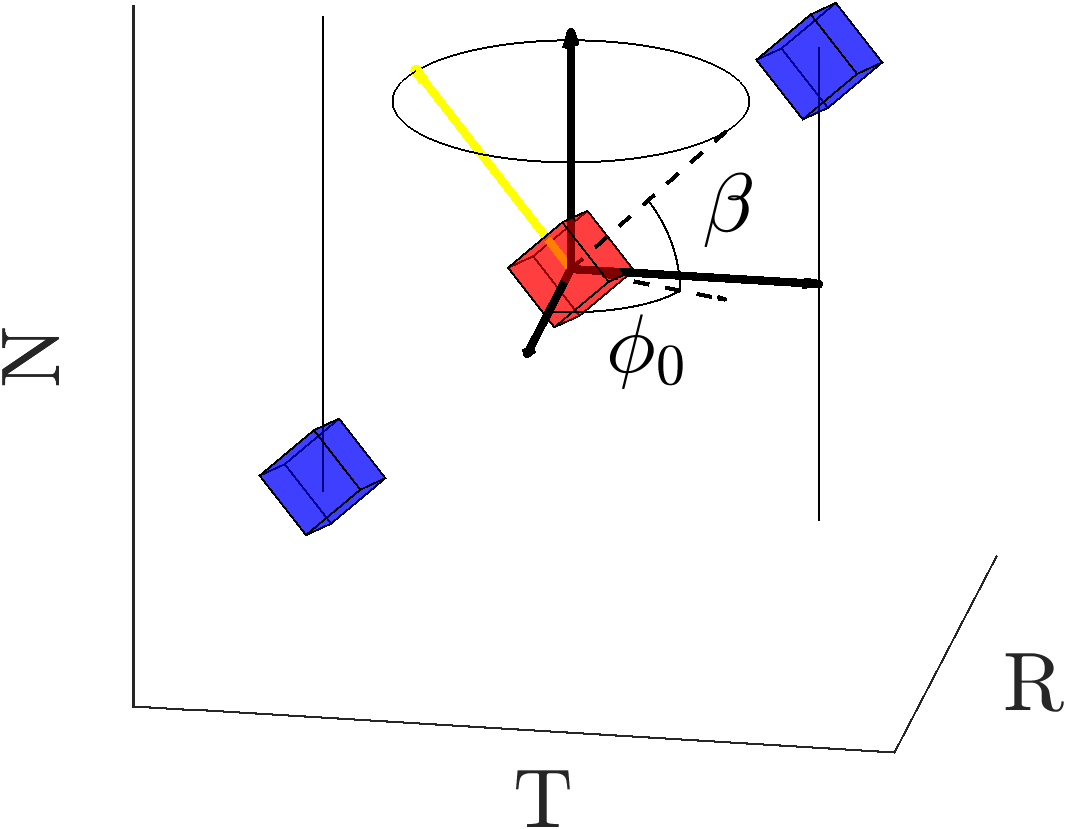}
        % \caption{1/2 Orbits}
    \end{subfigure}\hfill
    \begin{subfigure}{.23\textwidth}
        \centering
        \includegraphics[width=\linewidth]{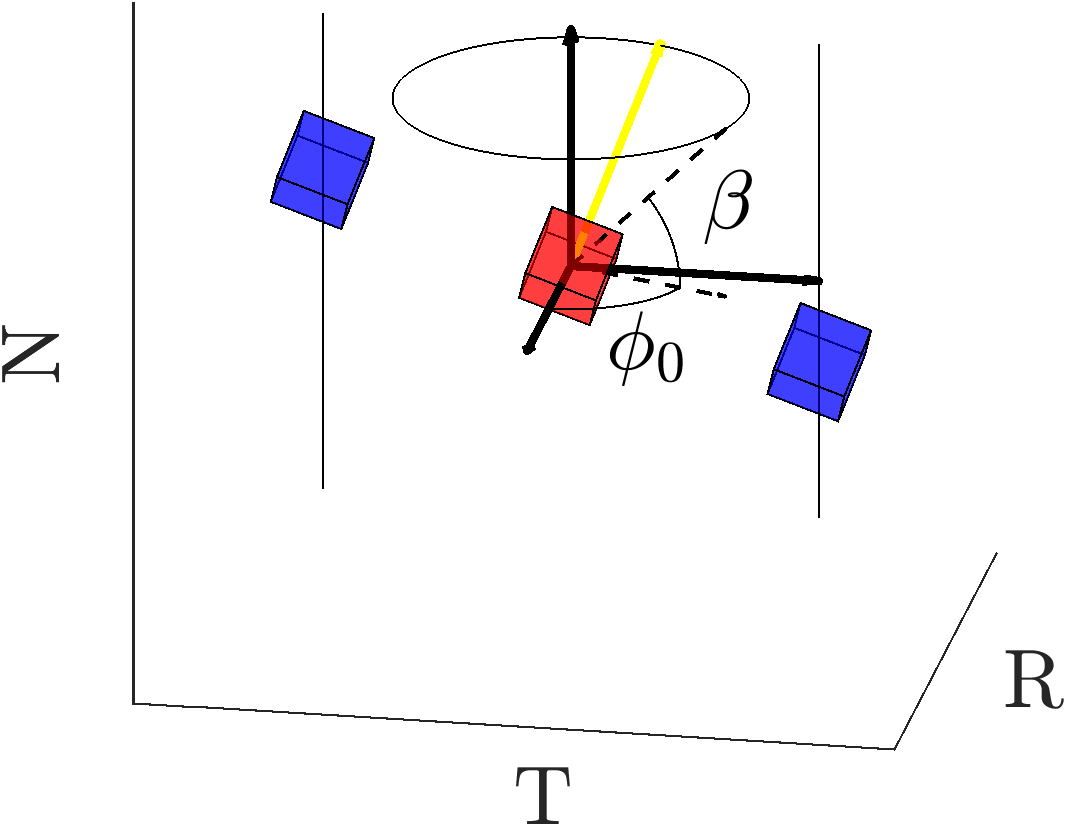}
        % \caption{3/4 Orbits}
    \end{subfigure}

    % \vspace{0.5em}

    % ---------- Row 2 ----------
    \begin{subfigure}{.23\textwidth}
        \centering
        \includegraphics[width=\linewidth]{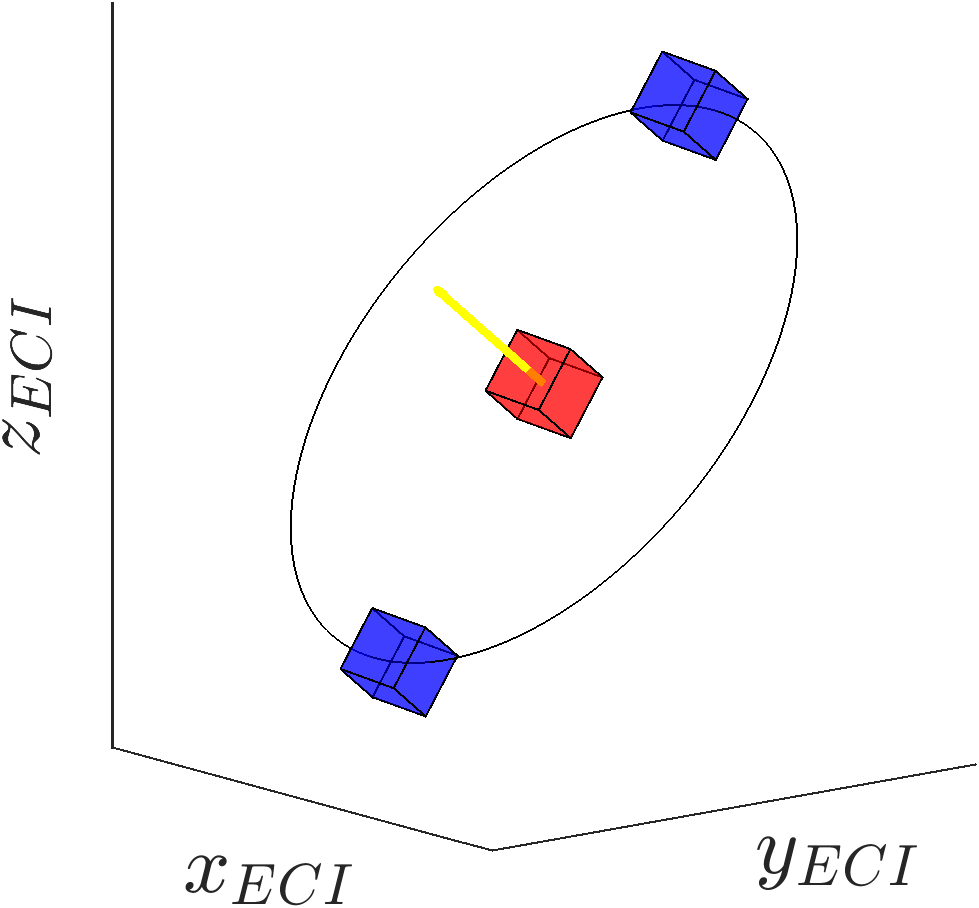}
        \caption{0 Orbits}
    \end{subfigure}\hfill
    \begin{subfigure}{.23\textwidth}
        \centering
        \includegraphics[width=\linewidth]{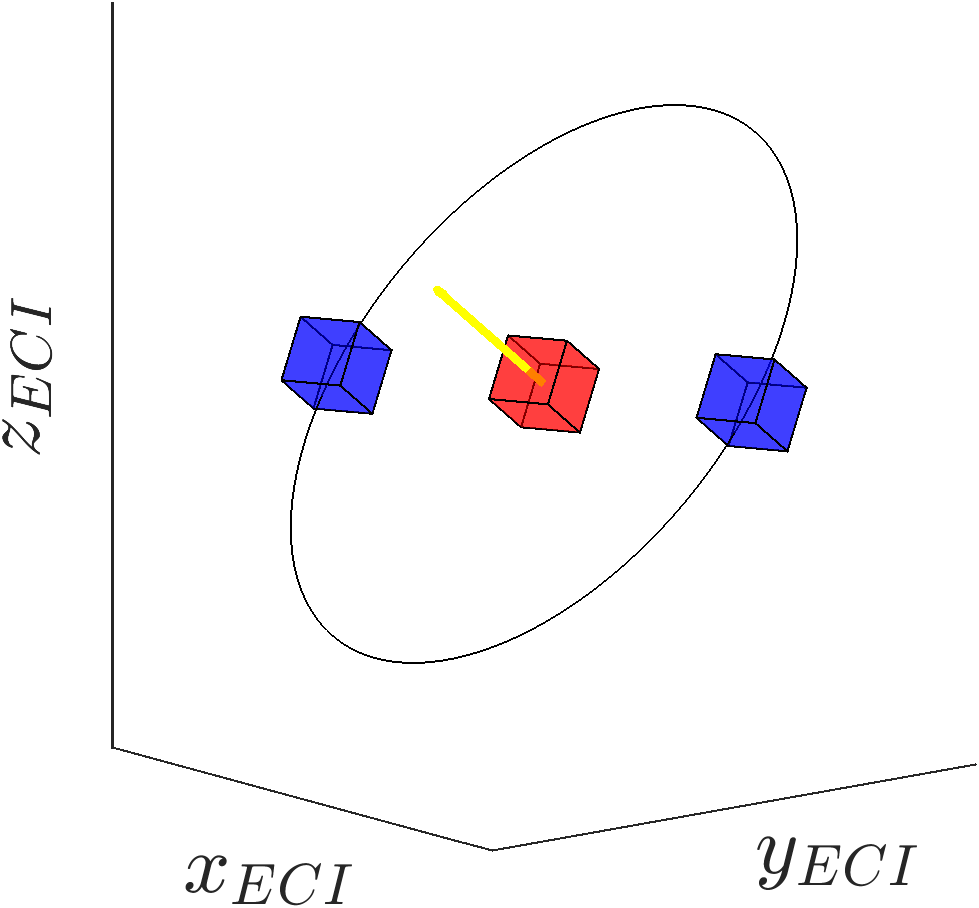}
        \caption{1/4 Orbits}
    \end{subfigure}\hfill
    \begin{subfigure}{.23\textwidth}
        \centering
        \includegraphics[width=\linewidth]{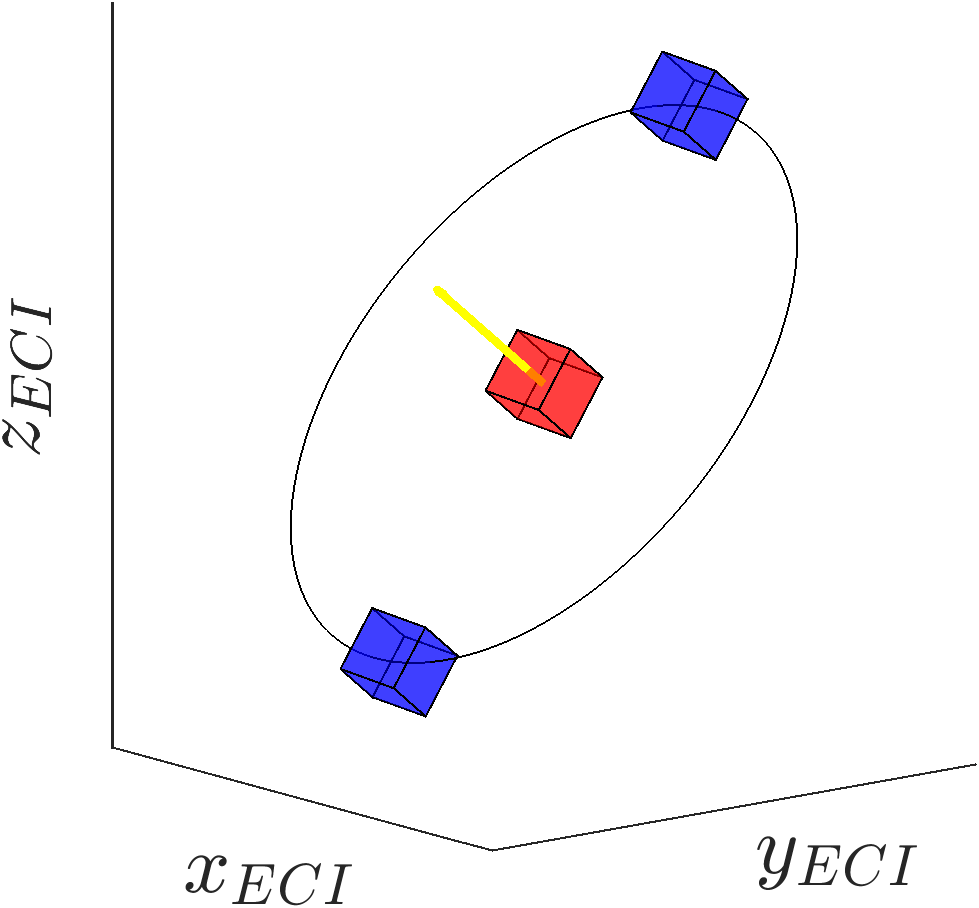}
        \caption{1/2 Orbits}
    \end{subfigure}\hfill
    \begin{subfigure}{.23\textwidth}
        \centering
        \includegraphics[width=\linewidth]{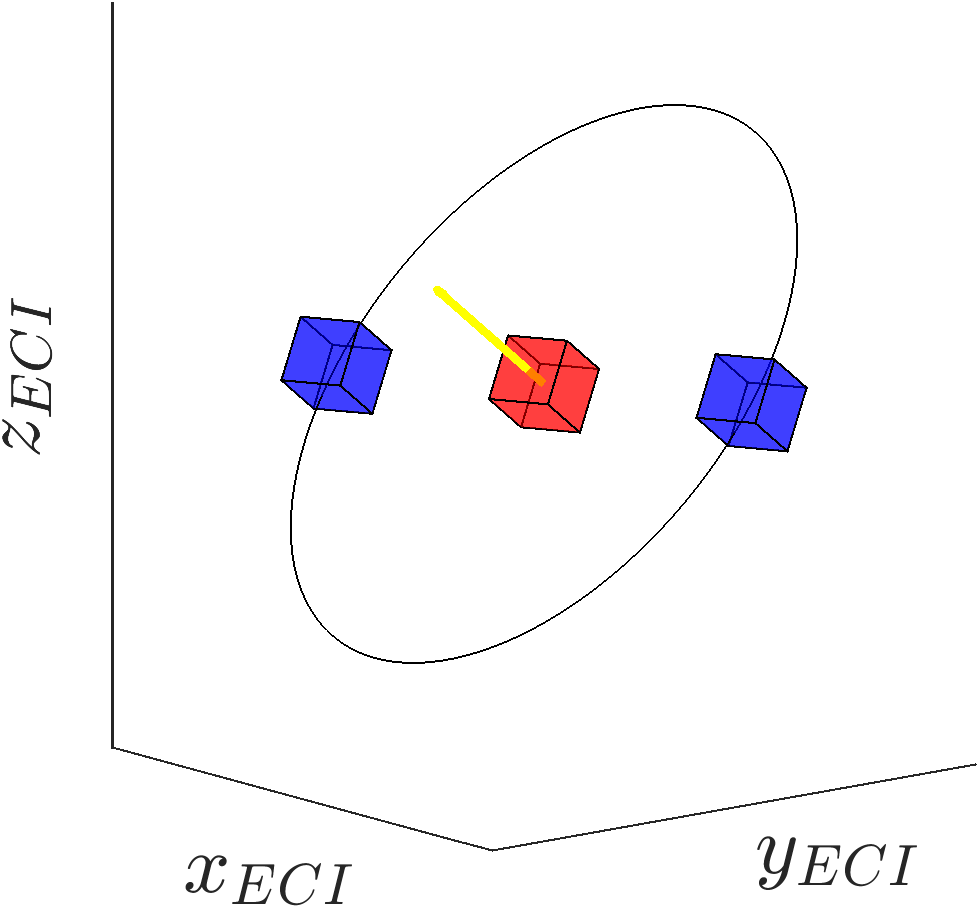}
        \caption{3/4 Orbits}
    \end{subfigure}

    \caption{Relative motion of a linear formation in the RTN and ECI frames over the course of one orbit. Note that the ratio of spacecraft size to inter-satellite distance is augmented for visual effect.}
    \label{fig:OrbitalVis}
\end{figure}

\begin{wrapfigure}{r}{0.4\textwidth}
\vspace{-.4in}
    \centering
    \begin{subfigure}{.18\textwidth}
        \centering
        \includegraphics[width=\linewidth]{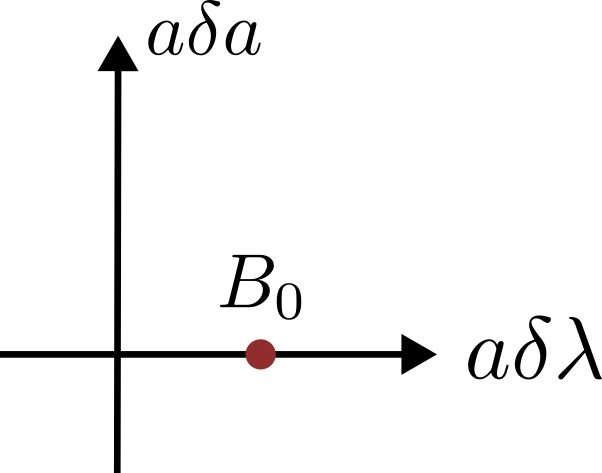}
        \caption{$\delta a- \delta \lambda$ plane}
    \end{subfigure} \hfill
    \begin{subfigure}{.19\textwidth}
        \centering
        \includegraphics[width=\linewidth]{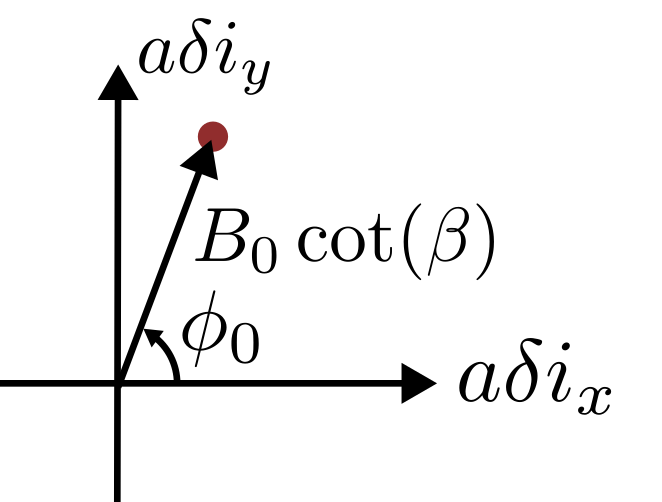}
        \caption{$\delta \bm{i}$ plane}
    \end{subfigure}
    \caption{Relative Orbital Elements associated with a linear formation}
    \label{fig:LinearROE}
    \vspace{-.2in}
\end{wrapfigure}

A visualization of the relative orbital motion of such a linear formation is displayed in Fig.~\ref{fig:OrbitalVis}. In the RTN frame, the spacecraft follow linear paths, with motion restricted to oscillation in the out-of-plane direction whose amplitude is $B_0/\tan\beta$ and constant offset in the along track direction. In the ECI frame, the two collector spacecraft rotate about the combiner spacecraft such that the relative orbital motion lies on a plane perpendicular to the observed star. Notably, the orbital motion expressed in relative orbital elements is stationary, and can be easily related to the $\beta$ and $\phi_0$ angles that describe the direction of observation with $\tan \phi_0 = \frac{\delta i_y}{\delta i_x}$ and $\cot \beta = \frac{a}{B_0} \| \delta \bm{i} \|_2$. This feature is depicted in Fig.~\ref{fig:LinearROE}, which shows the equivalent orbital motion in relative orbital elements space. 

\subsection{Concept of Operations}

An important property of the relative orbital motion of the aforementioned linear formations that demands consideration for devising a scheme to optimize observations is the lack of passive safety. Passively safe orbits can take the form of relative orbits that reduce collision risk by increasing RN plane spacecraft separation \cite{damico}. To account for the heightened risk produced by these scientific acquisition orbits, Ref.~\citenum{Rizza2026} proposes a nominal concept of operations in which these high stress science orbits are only inhabited for short durations, and passively safe orbits are inhabited at all other times. This allows for routine tasks, such as reaction wheel desaturation, downlinking, etc., to be performed in low-risk engineering orbits, while high accuracy navigation and control tasks are performed in higher-risk science orbits. Nominally, the spacecraft may reside in standby orbits which are parameterized by the relative orbital elements
\begin{equation}
    \delta \textit{\textbf{\oe}} = \pm \begin{bmatrix}
        0 & D_0/a & \delta e_x & \delta e_y & \delta i_x & \delta i_y
    \end{bmatrix}^{\top}
\end{equation}
\noindent where $D_0$ is an along-track separation parameter, on the order of hundreds of meters, and the relative eccentricity and inclination vectors are chosen such that $\delta \bm{e} || \delta \bm{i}$ and $a \min(\| \delta \bm{e} \|_2,\| \delta \bm{i} \|_2) \ge r_{min}$, which ensures passive safety by imposing a minimum RN separation of $r_{min}$ between any two spacecraft \cite{damico}. 
\begin{figure}[!htbp]
    \centering
    \includegraphics[width=1\linewidth]{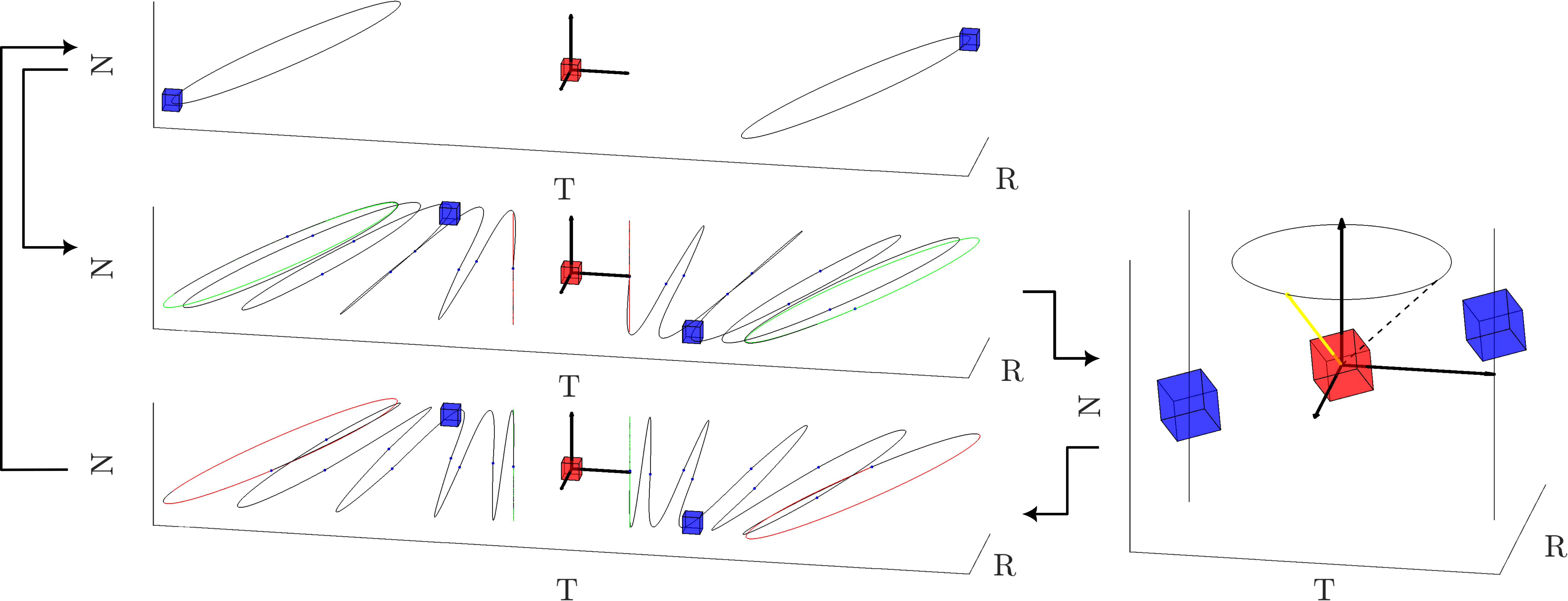}
    \caption{Example iteration of nominal concept of operations for an MIR mission. The formation resides in passively safe orbits, performs a transfer, resides in science orbits, then transfers back to standby. Note that the ratio of spacecraft size to inter-satellite distance is augmented for visual effect.}
    \label{fig:ConOps}
\end{figure}
Pursuing such a concept of operations requires transitioning between science and standby orbits during a science campaign, which poses constraints on both the frequency of scientific acquisition and the fuel required to perform a stellar observation. Fig.~\ref{fig:ConOps} demonstrates a singular science campaign performed under this concept of operations, whereby the formation performs a transfer from the initial standby orbit to the desired science orbit over the course of 5 orbits, performs relevant science tasks in the science orbit over the span of 5-10 orbits, and transfers back to a standby orbit over the span of 5 orbits. Such a science campaign would be performed over the course of about 1 day and at a frequency of about once per week, which is consistent with the relevant engineering tasks that need to be performed between observations \cite{Rizza2026}.

\section{Observability Constraints}

% Though the linear formation relative orbit design ensures the capability of achieving continuous perpendicularity of the linear formation with respect to an inertial direction, several constraints on a mission, spacecraft, and orbit are relevant for determining which inertial directions can be reasonably observed and the duty cycle at which they can be maintained. Such constraints inform the set of stars that can be observed and for what durations. Particularly, these constraints stem from Sun lighting, Earth occlusion, and relative orbit oscillation effects \cite{Hansen_2020}.
The linear formation relative orbit design ensures theoretically continuous perpendicularity with respect to a star direction for small spacecraft separations. However, a given star may not be observable continuously due to relevant mission, spacecraft, and orbital constraints. To inform the set of stars that may be observed and for what durations, this section establishes the relevant observability constraints and their effect on scientific acquisition. Any given star of the celestial sphere is determined by a right ascension $\alpha$ and declination $\delta$, which maps to a unit vector direction from the origin of the Earth by $\hat{\bm{s}}^\top = \begin{bmatrix}
    \cos(\alpha)\cos(\delta) \ \sin(\alpha) \cos(\delta) \ \sin(\delta)
\end{bmatrix}$, which is expressed in the celestial equatorial plane with the first basis pointing towards vernal equinox. This paper considers only Sun-fixed orbits, that is, Sun-synchronous orbits, which have year-long and consistent observability of the celestial sphere.

\begin{figure}[!tbp]
    \centering

    % ---------- Row 1 ----------
    \begin{subfigure}{.3\textwidth}
        \centering
        \includegraphics[width=\linewidth]{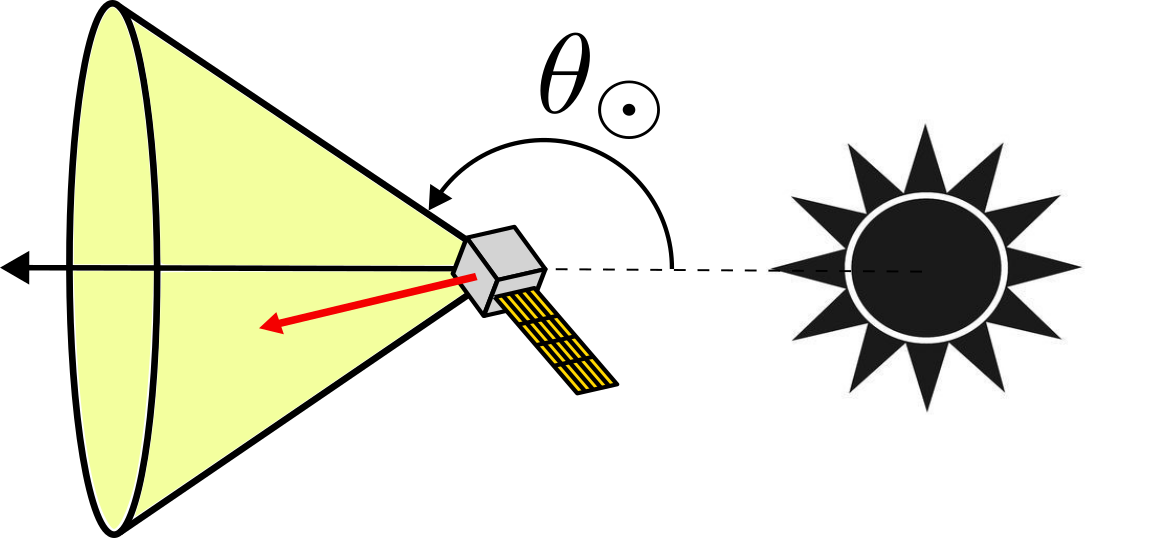}
        \caption{Sun Exclusion}
    \end{subfigure}\hfill
    \begin{subfigure}{.3\textwidth}
        \centering
        \includegraphics[width=\linewidth]{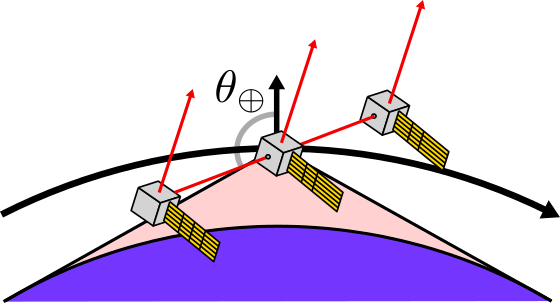}
        \caption{Earth Occlusion}
    \end{subfigure}\hfill
    \begin{subfigure}{.3\textwidth}
        \centering
        \includegraphics[width=\linewidth]{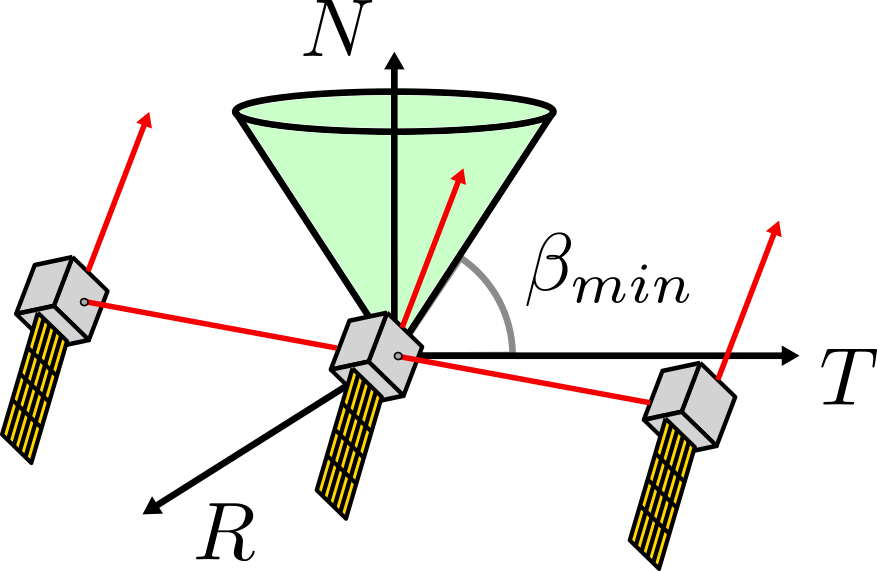}
        \caption{Baseline Oscillation}
    \end{subfigure}\hfill
    \caption{Conceptualization of the Sun exclusion, Earth occlusion, and relative motion constraints on field of regard. The intersection of these regions defines the visible regions of the celestial sphere.}
    \label{fig:ViewingConstraints}
\end{figure}

\subsection{Sun Exclusion}

Sun exclusion is a dominant pointing constraint on observing extrasolar systems. For many space telescope missions, the accessible viewing regions of the celestial sphere, known as the field of regard (FOR), are dominated by Sun exclusion, which is enforced to protect scientific instruments from thermal heating or stray light. For example, missions like JWST \cite{jwst_mocd_2011} have a field of regard between $85\degree \le \theta_{\odot} \le 135 \degree$, where $\theta_{\odot}$ is the anti-solar angle, defining an annular region that observes the poles of the celestial sphere. This allows for global accessibility of the celestial sphere over the span of a year. In contrast, LEO space interferometers have a Sun exclusion zone defined simply by $\theta_{\odot} \ge \theta_{\odot,min}$, defining a conical region. This constraint is driven by the need to avoid stray light impinging the scientific devices \cite{Hansen_2020}. Typical studies consider $\theta_{\odot,min} = 120\degree-135\degree$ \cite{Hansen_2020}. The Sun exclusion constraint can be expressed mathematically as 
\begin{equation}
    g_{\odot}(\alpha,\delta,\lambda_\odot,i,LTAN) = 
    -\hat{\bm{s}}_{\odot}^{\top} \hat{\bm{s}} +\cos(\theta_{\odot,min}) \ge 0,
\end{equation}
% \begin{equation}
%     \bm{r}_{sc/\odot}^{\top} \hat{\bm{s}} \ge \cos(\theta_{\odot,max}) \iff g_{\odot}(\alpha,\delta,\Omega(t),i,t) \le0,
% \end{equation}
\noindent where $\hat{\bm{s}}_{\odot}^\top = [\cos\lambda_\odot \ \sin \lambda_\odot \cos \epsilon \ \sin \lambda_\odot \sin \epsilon]$ is the unit vector from the Earth to the Sun in the celestial equatorial frame and $\hat{\bm{s}}$ is the unit vector observation direction, which are dependent on the observation angles ($\alpha$ and $\delta$) as well as the solar ecliptic longitude ($\lambda_\odot \approx \frac{2 \pi}{365.2422}\left(t-t_{\text {vernal }}\right)$), solar obliquity ($\epsilon = 23.44\deg$), orbital inclination $i$, and local time of the ascending node $LTAN$.

\subsection{Earth Occultation}

For certain observation directions, the Earth may occult the star for portions of the absolute orbit. The fraction of the orbit for which an inertial direction is visible is termed the visibility fraction and is computed with
\begin{equation}
    f_{vis,\oplus}(\alpha,\delta,\lambda_\odot,i) = \frac{1}{\pi} \Re \left(\arccos\left( \frac{\cos(\theta_{\oplus})}{\cos(\beta)}\right)\right).
\end{equation}
Here, $\theta_{\oplus} \ge 90 \degree$ describes the Earth surface angle from the radial direction of the orbit, which may depend on the orbital altitude, and $\Re()$ denotes the real component returned by $\arccos$. This study considers $\theta_{\oplus} = \pi/2+\arccos(R_{\oplus}/a)$.

\subsection{Solar Occultation} 

It is possible for observation of a point on the celestial sphere to be visible during the times at which the spacecraft are in the Earth's shadow, in which case Sun exclusion no longer constrains observation. To analyze these cases, the movement of the Sun pointing vector in the RTN frame can be defined as $\hat{\bm{s}}_{\odot,RTN}^\top = \begin{bmatrix}
    \cos(\beta_{\odot}) \cos(\phi_\odot-u) & \cos(\beta_\odot) \sin(\phi_{\odot}-u) & \sin(\beta_\odot)
\end{bmatrix}$ over short durations. Determining the portion of the orbit for which a given star is visible in the case of solar occultation amounts to determining $\mathcal{U}$, the range of $u$ for which both the Sun is occulted and the star is not occulted by the Earth, i.e., 
\begin{equation}\mathcal{U} = \{u \in [0,2\pi] | \hat{\bm{e}}_1 \cdot \hat{\bm{s}}_{RTN} \ge \cos \theta_\oplus\} 
\bigcap \{u \in [0,2\pi] | \hat{\bm{e}}_1 \cdot \hat{\bm{s}}_{\odot,RTN} \le \cos \theta_\oplus\},\end{equation}
\noindent where $\hat{\bm{e}}_1^\top = [1 \ 0 \ 0]$. The visibility fraction can then be computed from the length of this interval, $f_{vis,\odot}(\alpha,\delta,\lambda_\odot,i) = |\mathcal{U}|/2\pi$, via the expression
\begin{equation}
    f_{vis,\odot}(\alpha,\delta,\lambda_\odot,i) = \frac{1}{2\pi}\begin{cases}
        0, & u_\oplus + u_\odot \le \Delta \phi \\
        u_\oplus + u_\odot - \Delta \phi & |u_\oplus - u_\odot | \le \Delta \phi \le u_\oplus + u_\odot \\
          2\min\left(u_\oplus,u_\odot\right) & \Delta \phi \le |u_\oplus - u_\odot |
    \end{cases}
    \label{eq:SolarOccultation}
\end{equation}
\noindent where $u_\oplus = \Re \left(\arccos\left( \frac{\cos \theta_\oplus}{\cos \beta} \right) \right)$, $u_\odot = \pi - \Re\left(\arccos\left( \frac{\cos \theta_\oplus}{\cos \beta_\odot} \right)\right)$, and $\Delta \phi = \arccos(\cos(\phi_0-\phi_\odot))$.

\subsection{Baseline Oscillation}

Nulling interferometers typically impose a constraint on the maximum oscillation of the inter-satellite distance for scientific purposes. This quantity corresponds to $\cot(\beta)$, which maps to a minimum observation angle $\beta_{min}$. Accordingly, the baseline oscillation constraint can be written as
\begin{equation}
    g_{\beta}(\alpha,\delta,\lambda_\odot,i) =|\beta|-\beta_{min} \ge 0.
\end{equation}

\subsection{Frozen Relative Orbit Constraint} 

For mission design of formation flight missions, it is common to enforce that any relative orbit to be inhabited be frozen in the presence of dominant perturbations. In polar, low-Earth orbits, this dominant perturbation comes in the form of the $J_2$ harmonic, for which the linear formation relative orbit remains frozen in science orbits whenever $\delta i_x \approx 0$ \cite{Rizza2026}. This corresponds to observations for which $\phi_0 = \pi/2$ or $\phi_0 = 3\pi/2$. As shown in Ref.~\citenum{Rizza2026}, this reduces the percentage of time during the year for which a given target is visible but does not reduce the number of targets that are available over a full year. Such a constraint can be convenient to enforce from a mission design perspective and results in less fuel expenditure to maintain science orbits. This constraint enforces that observations of a star, with associated $(\alpha,\delta)$, occur precisely when
\begin{equation}
    g_{\delta i}(\alpha,\delta,\lambda_\odot,LTAN) = (\lambda_\odot- \alpha + LTAN + \pi/2) \mod (2\pi) = 0,
    \label{eq:diyconstraint}
\end{equation}
\noindent meaning a given star can only be observed on one or two occasions per year, dependent upon other observability constraints.

To summarize, the full duty cycle of observations, that is, the portion of a single orbit for which a stellar target is visible, is given by
\begin{equation}
    f_{vis}(\alpha,\delta,\lambda_\odot,i,LTAN) = (1_{g_\odot \ge 0} \cdot f_{vis,\oplus} + 1_{g_\odot < 0} \cdot f_{vis,\odot}) \cdot 1_{g_\beta \ge 0} \cdot 1_{g_{\delta i} =0}
    \label{eq:fvis1}
\end{equation}
when the frozen relative orbit constraint is active, and 
\begin{equation}
    f_{vis}(\alpha,\delta,\lambda_\odot,i,LTAN) = (1_{g_\odot \ge 0} \cdot f_{vis,\oplus} + 1_{g_\odot < 0} \cdot f_{vis,\odot}) \cdot 1_{g_\beta \ge 0}
    \label{eq:fvis2}
\end{equation}
otherwise.

\subsection{Fuel Expenditure} % Include delta i_x constraint here
A space interferometry mission is naturally constrained by the amount of fuel it can expend over the mission duration. Fuel expenditure in these contexts is primarily driven by the operational transfers between science and standby orbits and station keeping of science orbits \cite{Rizza2026}. As a result, the scientific acquisition and fuel expenditure are fundamentally linked. 

To quantify the fuel expenditure necessary to transition between two stars, analytical bounds on fuel expenditure based on relative orbital elements can be leveraged. In particular, the expression
\begin{equation}
    \Delta V_{lb} = an \left(\max\left\{ \frac{|\Delta \delta a|}{2}, \frac{|\Delta \delta \lambda |}{2\Delta M},\frac{\| \Delta \delta \bm{e}\|}{2} \right\} + \| \Delta \delta \bm{i} \| \right)
\end{equation}
\noindent gives a lower bound on $\Delta V$ expenditure given desired relative orbital element differences $\Delta \delta a$, $\Delta \delta \lambda$, $\Delta \delta \bm{e}$, and $\Delta \delta \bm{i}$ \cite{chernickphd}. However, the transfers relevant for LEO space interferometry reduce to 
\begin{equation}
    \Delta V = an (\| \Delta \delta \bm{e} \|/2 + \| \Delta \delta \bm{i} \| )
\end{equation} 
\noindent since $\Delta \delta a = 0$ and $\Delta M$ is large. Numerical results in Ref.~\citenum{Rizza2026} have shown this analytical expression to be accurate for these scenarios. 

To determine the amount of fuel expenditure necessary to transition between two observable star directions, i.e., from $(\alpha_0,\delta_0)$ to $(\alpha_f,\delta_f)$, it must be taken into account that the formation must undergo a standby phase between observations. In this standby phase, RN separation must be enforced, so that the relative orbital elements of the formation satisfy $a\| \delta \bm{e}_{sb} \| \ge r_{min}$ and $a\| \delta \bm{i}_{sb} \| \ge r_{min}$. Frozenness of the standby orbit can also be enforced by setting $\delta e_{x,sb} = \delta i_{x,sb} = 0$. The star observation directions map uniquely to points on the inclination plane with $\delta \bm{i} = B_0 [ \cos\phi_0 \cot \beta, \sin \phi_0 \cot \beta]^{\top}$. Therefore, the $\Delta V$ cost of transitioning between two observations whose associated locations on the relative inclination plane are $\delta \bm{i}_0$ and $\delta \bm{i}_f$ is given by
\begin{equation}
\begin{aligned}
    \Delta V_{0 \rightarrow f} &= an \left( \min_{a \min\{ | \delta i_{y,sb}|,|\delta e_{y,sb} | \} \ge r_{min}} \left\|\delta \bm{i}_0- \begin{bmatrix} 0 \\ \delta i_{y,sb} \end{bmatrix} \right\| + \left\| \delta \bm{i}_f - \begin{bmatrix} 0 \\ \delta i_{y,sb} \end{bmatrix} \right\| + | \delta e_{y,sb} | \right) .
    % \\
    % &= n \left( r_{min} + a\min_{a \| \delta \bm{i}_{sb}\|\ge r_{min}} \|\delta \bm{i}_0-\delta \bm{i}_{sb} \| + \| \delta \bm{i}_f - \delta \bm{i}_{sb} \|\right)
    \end{aligned}
\end{equation}
The minimizing $\delta e_{y,sb}$ and $\delta i_{y,sb}$ arguments are decoupled. For the former, the minimizing argument is evidently $|\delta e_{sb}| = r_{min}$. The latter consists of minimizing the distance between two points with an intersection of the $\delta i_x$ axis. This minimizer has the analytical solution
\begin{equation}
    \delta i_{y,sb} = \begin{cases}
        \max\{ \delta i^*_{y,sb},r_{min}/a\} \ & \text{if} \ \delta i^*_{y,sb} \ge 0 \\
        \min\{ \delta i^*_{y,sb},-r_{min}/a\} \ & \text{otherwise}
    \end{cases}, 
\end{equation}
where
\begin{equation}
    \delta i^*_{y,sb} = \begin{cases} \frac{\delta i_{y,f} \delta i_{x,0} + \delta i_{y,0} \delta i_{x,f}}{\delta i_{x,0} + \delta i_{x,f}} \ & \text{if}  \ \delta i_{x,0}+\delta i_{x,f} \neq 0 \\ \frac{\delta i_{y,f} + \delta i_{y,0}}{2} \ &\text{otherwise} \end{cases}.
\end{equation}
A generalization of this result to the full $\delta \bm{i}$ plane is discussed in \nameref{app:Transfer} but is not considered in the presented results.

To measure $\Delta V$ expenditure for science station keeping , this work leverages an empirically derived analytical expression of the form 
\begin{equation}
    \Delta V_{sk} = \frac{t_{eff,max}}{t_{orbit}}\sqrt{\frac{a_0^7}{a^7}} \left( \Delta V_{\delta \lambda} |a \delta \lambda| + \Delta V_{\delta i_x} |a \delta i_x| + \Delta V_{\delta i_y} |a \delta i_y| \right),
    \label{eq:SK}
\end{equation}
where $\Delta V_{\delta \lambda}$, $\Delta V_{\delta i_x}$, and $\Delta V_{\delta i_y}$ are numerical terms that quantify fuel expenditure as weighted by the associated relative orbital elements of the science orbit. The nominal science orbit relative orbital elements are given in Eq.~\ref{eq:scienceroe} are a function of the nominal observation baseline $B_0$, which is chosen as a function of spectral properties of the stellar target, as will be discussed in the \nameref{sec:Science} section, along with $\phi_0$ and $\beta$, which are uniquely determined by $(s,\lambda_\odot)$. Station keeping fuel expenditure increases linearly with increasing $\delta \lambda$, $\delta i_x$, and $\delta i_y$ as a result of increasing secular and osculating motion effects that must be suppressed to keep the optical path difference $\Lambda$ (Eq.~\ref{eq:OPD}) low. Furthermore, these effects diminish with increasing altitude at a rate of $a^{-7/2}$. This study considers $\Delta V_{\delta \lambda}  = 0.5 \ \frac{\textrm{mm/s}}{\textrm{m orbit}}$, $\Delta V_{\delta i_x} = 0.5 \ \frac{\textrm{mm/s}}{\textrm{m orbit}}$, and $\Delta V_{\delta i_y} = 1.5 \ \frac{\textrm{mm/s}}{\textrm{m orbit}}$, calibrated at $a_0 = 6878 \ \textrm{km}$.

To summarize, the $\Delta V$ expenditure to transfer from observation $(s_0, \lambda_{\odot,0})$ to $(s_f, \lambda_{\odot,f})$ consists of transfering to a new science orbit, followed by station keeping the science orbit, giving the final expression
\begin{equation}
    \Delta V(s,\lambda_{\odot},s^{\prime},\lambda_{\odot}^\prime) = \Delta V_{s \to s^{\prime}} + f_{vis}(s^{\prime},\lambda_{\odot}^{\prime})\Delta V_{sk}(s^{\prime},
    \lambda_\odot^\prime) .
    \label{eq:DV}
\end{equation}
Station keeping cost is weighted by visibility fraction since station keeping only needs to be performed during observation times.

\subsection{Operational Dwell}

The final consideration relates to operational dwell between observations. In practice, observations are to be performed on a basis that is consistent with engineering demands in standby orbits. For this study, an operational dwell range of between one and two weeks between observations is enforced, which corresponds to $\Delta \lambda_{\odot,min} = 2\pi \frac{7}{365.24}$ and $\Delta \lambda_{\odot,max} = 2\pi \frac{14}{365.24}$.

\section{Scientific Ranking and Baseline Selection} \label{sec:Science}

Optimizing the schedule of scientific observations requires metrics for the reward of observing a given star. This paper measures the scientific potential of a given stellar system via \textit{yield}, a commonly used metric in exoplanet detection \cite{starkMAXIMIZINGExoEarthCANDIDATE2014}. Yield describes the expected scientific return, which in this case means expected number of detected exoplanets. It is estimated as
\begin{equation}
    Y_i =  \eta_{\oplus,i} C_i
\end{equation}
\noindent where $\eta_{\oplus,i}$ estimates the expected number of exoplanets in the $i$th stellar system and $C_i$ estimates the \textit{completeness} of the $i$th stellar system, which is the probability of an exoplanet detection associated with a given system, given observation parameters. 

To determine the yield associated with a given system, it is important to introduce the signal-to-noise ratio which is often used in nulling interferometry missions to establish baselines for a measurement success \cite{dannertLargeInterferometerExoplanets2022}. This work uses a simplified SNR metric in comparison to the one proposed in Ref.~\citenum{dannertLargeInterferometerExoplanets2022}, which is determined for a Bracewell configuration in Sun-Earth $L_2$, for the specific linear interferometer of interest, with the form
\begin{equation}
    SNR = \sqrt{ \frac{t_{eff}}{2}} \frac{R_{sig}}{\sqrt{R_{leak} + R_{zodi} + R_{exo}}}.
    \label{eq:SNR}
\end{equation}
\begin{wrapfigure}{r}{2.2in}
\vspace{0in}
    \centering
        \includegraphics[width=2in]{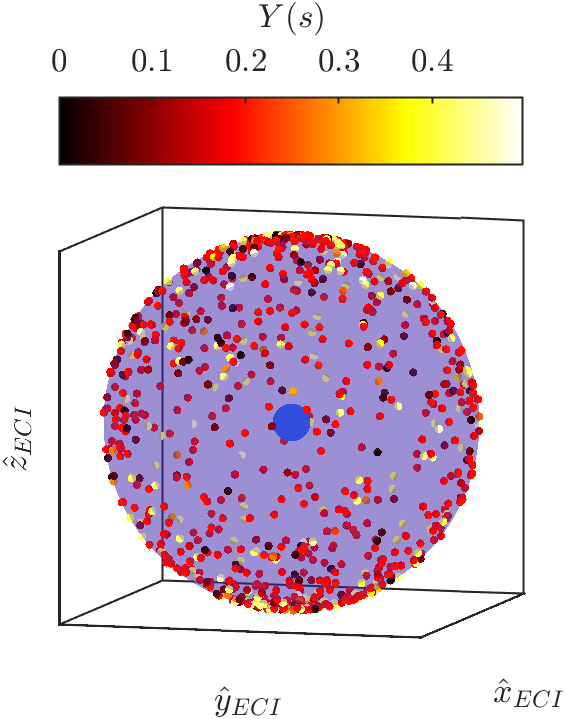}
    \caption{Stellar targets colored by associated yield $Y(s)$ projected onto the celestial sphere.}
    \label{fig:StarYield}
    \vspace{-.2in}
\end{wrapfigure}
\noindent Here, $t_{eff}$ is the integration time in seconds, i.e., the time duration for which the system is observed, $R_{sig}$ is the magnitude of the planet signal in photons per second, $R_{leak}$ is the magnitude of the stellar leakage signal (which is suppressed by the nulling interferometer) in photons per second, and $R_{zodi}$ and $R_{exo}$ are the magnitudes of local zodiacal and exozodiacal dust noises in photons per second. Ref.~\cite{dannertLargeInterferometerExoplanets2022} establishes a baseline signal-to-noise ratio of $SNR = 7$ for successful detection. To determine completeness from this metric, this paper establishes a maximum integration time $t_{eff,max}$ and performs Monte Carlo sampling of various exoplanet orbital parameters within the habitable zone of the system, and determines which samples exceed the $SNR=7$ threshold within the maximum integration time. Concretely,
\begin{equation}
    C_i = \frac{1}{N_{mc}} \sum_{j=1}^{N_{mc}} 1_{\{t_{eff,j} \le t_{eff,max}\}}.
    \label{eq:Completeness}
\end{equation}
where $N_{mc}$ is the number of Monte Carlo sammples, $t_{eff,j}$ is the observation time required for detection of sample $j$, and $t_{eff,max}$ is the maximum allowable observation duration, which is set to 1 day (Tab.~\ref{tab:ScienceParameters}). For further details on the computation to determine whether a given sample is detectable, see \nameref{app:Science}. 

Stellar target data is determined from the Habitable Worlds Observatory stellar target catalog \cite{Tuchow_2025}, which contains data for over 10,000 stellar targets of interest for exoplanet detection. Parameters used for yield computation can be found in Tab.~\ref{tab:ScienceParameters}. Stellar targets that have a yield of zero, i.e., $Y_i = 0$, are removed, resulting in a total of 935 targets to optimize over. The resulting targets are displayed in Fig.~\ref{fig:StarYield}.

In addition, the spectral properties of the system inform the baseline $B_0$ that maximizes scientific potential. The optimal baseline of the linear array is chosen such that 
\begin{equation} B_0 = 0.59 \frac{\lambda_0}{\theta_{HZ}}, \end{equation} 
where $\lambda_0$ is the central wavelength to measure and $\theta_{HZ}$ is the angular separation associated with the habitable zone, consistent with Ref.~\citenum{dannertLargeInterferometerExoplanets2022}. This value is also clipped such that $B_0 \in [B_{min},B_{max}]$.

\section{Observation Scheduling Optimization}

Given the relevant constraints on observation established in the previous section, the observation scheduling problem can now be posed in optimization form. Given a catalog of stars expressed by the list of quadruplets $S=\{s_k\}_{k=1}^N = \{ (\alpha,\delta,Y,B_0)_k \}_{k=1}^{N}$, which consists of locations on the celestial sphere in terms of right ascension $\alpha$ and declination $\delta$, the encoded scientific reward/yield $Y_k \in [0,1]$, and the nominal baseline distance $B_0$ from which to measure the stellar target, an optimization problem must determine a sequence of stellar observations and the times of the year at which they must be performed, encoded by the ordered path $P = \{(s,\lambda_\odot)_1,(s,\lambda_\odot)_2, \cdots,(s,\lambda_\odot)_n\}$ that maximizes scientific outcome while satisfying observability constraints. This optimization problem can be stated as
% \begin{equation}
%     \begin{aligned}
%         \max_{P \in \mathcal{P}} \quad &\sum_{(\alpha,\beta,R,t)_j\in P} R_j f_{vis}(\alpha_j,\beta_j,\Omega(t_j),i(t_j)) \\
%         \text{s.t.}\quad & g_{\odot}(\alpha_j,\delta_j,\Omega(t_j),i(t_j),LTAN) \ge 0 \ \forall (\alpha_j,\beta_j,R_j,t_j)\in P \\
%         & g_{\beta}(\alpha_j,\delta_j,\Omega(t_j),i(t_j)) \ge 0 \ \forall (\alpha_j,\beta_j,R_j,t_j)\in P \\
%         & \sum_{(j,j+1) \in P} \Delta V((\alpha,\beta,t)_j,(\alpha,\beta,t)_{j+1}) \le \Delta V_{max} \\
%         & t_j + \Delta t_{min} \le t_{j+1} \ \forall (j,j+1) \in P
%     \end{aligned}
%     \label{eq:OptimizationProblem}
% \end{equation}
\begin{equation}
    \begin{aligned}
        \max_{P} \quad &\sum_{(s,\lambda_\odot)_j\in P} Y_j f_{vis}(s_j,\lambda_{\odot,j}) \\
        \text{s.t.}\quad &  \sum_{(s,\lambda_\odot)_{(j,j+1)} \in P} \Delta V(s_j,\lambda_{\odot,j},s_{j+1},\lambda_{\odot,j+1}) \le \Delta V_{max} \\
        & \lambda_{\odot,j+1}-\lambda_{\odot,j} \in[\Delta \lambda_{\odot,min},\Delta \lambda_{\odot,max}] \ \forall (s,\lambda_\odot)_{(j,j+1)} \in P
    \end{aligned}
    \label{eq:OptimizationProblem}
\end{equation}
Here, $(s,\lambda_\odot)_j\in P$ denotes an element of the path and $(s,\lambda_\odot)_{(j,j+1)}$ denotes a two element sequence of the path. Recall that $f_{vis}(s_j,\lambda_{\odot,j})$ gives the observability fraction of star $s_j$ at time $\lambda_{\odot,j}$, as computed from Eq.~\ref{eq:fvis1} or Eq.~\ref{eq:fvis2}, and that $\Delta V(s_j,\lambda_{\odot,j},s_{j+1},\lambda_{\odot,j+1})$ gives the $\Delta V$ expenditure associated with transition from observation $(s,\lambda_\odot)_j$ to $(s,\lambda_\odot)_{j+1}$, as computed from Eq.~\ref{eq:DV}. Furthermore, $\Delta \lambda_{\odot,min}$ and $\Delta \lambda_{\odot,max}$ represent operational dwell constraints along the path. The reward incorporates observability given by Eqs.~\ref{eq:fvis1} or \ref{eq:fvis2} via a weighting on the scientific yield, which naturally enforces that any given path contain only visible stars for some fraction of time. Stellar targets which are never visible over the mission duration, i.e., $\{ s \in S | f_{vis}(s,\lambda_\odot) =0 \ \forall \lambda_\odot\}$ can immediately be removed from the set of optimization targets, e.g., polar targets for a Sun-synchronous terminator orbit. 

The form of the optimization problem incorporates Sun exclusion, Earth occultation, solar occultation, baseline oscillation, fuel expenditure and operational dwell constraints, however it is flexible in the incorporation of the frozen relative orbit constraint. In the case that this constraint is active, the problem can be further reduced by the fact that the time at which a star is observed, $\lambda_{\odot,j}$, is now constrained such that $\lambda_{\odot,j} = \alpha_j + LTAN \pm \pi/2$.

\subsection{Solution Approach}

\begin{figure}
    \centering
    \includegraphics[width=0.6\linewidth]{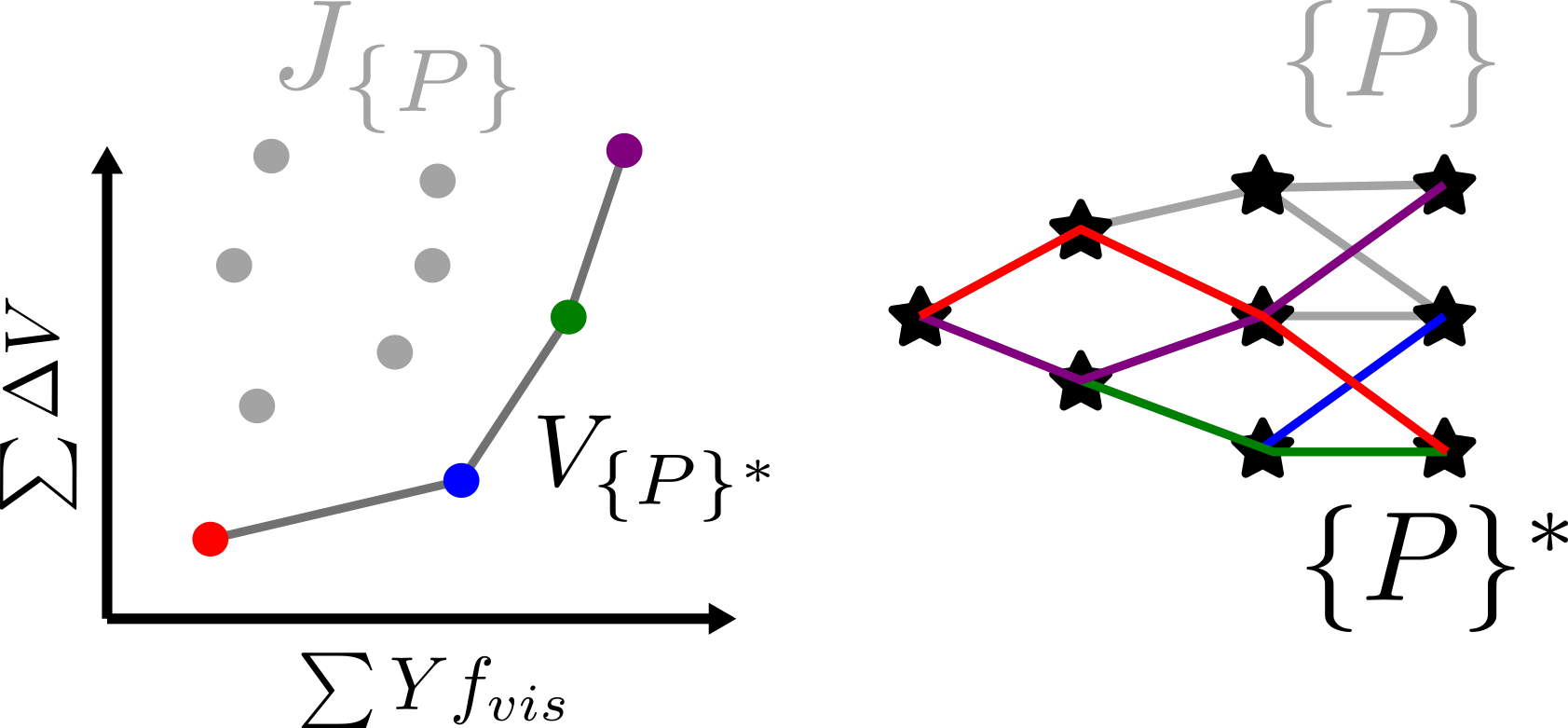}
    \caption{Conceptualization of the multi-objective dynamic programming definitions. Given some possible ensuing feasible paths $\{P\}$, which have an associated scientific reward and fuel expenditure $J_{\{P\}}$, the value function $V_{\{P\}^*}$ is defined as the set of Pareto-dominant objectives and is associated with the optimal paths $\{P\}^*$.}
    \label{fig:pareto}
\end{figure}

The optimization problem posed in Eq.~\ref{eq:OptimizationProblem} can be considered an orienteering problem \cite{VANSTEENWEGEN20111}. Such problems describe the optimization of a path to maximize the cumulative reward of the visited nodes subject to a constraint on the length of the path. In Eq.~\ref{eq:OptimizationProblem}, each star represents a node with associated reward and the constraint on fuel expenditure represents a constraint on the path traveled.

This work proposes a solution approach based on multi-objective dynamic programming, whereby both the fuel and scientific reward are simultaneously optimized and non-optimal solutions are pruned via a Pareto front. This technique provides a tractable strategy for solving Eq.~\ref{eq:OptimizationProblem} to global optimality and provides a global policy from any initial condition, making the strategy highly desirable for a mission which may involve varying fuel expenditure and the desire to perform revisits of stellar targets. The weighting of scientific reward to fuel expenditure can be encoded by the multi-objective
\begin{equation}
    J_{\{P\}}(s_0,\lambda_{\odot,0}) = \left\{\begin{matrix}
        \sum_{(s,\lambda_\odot)_j \in P} Y_j f_{vis}(s_j,\lambda_{\odot,j}) \\
        -\sum_{(s,\lambda_\odot)_{(j,j+1)} \in P} \Delta V(s_j,\lambda_{\odot,j},s_{j+1},\lambda_{\odot,j+1}) 
    \end{matrix}\right\},
\end{equation}
which consists of the discrete list of scientific rewards and $\Delta V$ expenditures associated with the set of discrete feasible paths $\{P\}$ starting from $(s_0,\lambda_{\odot,0})$, which are defined as 
\begin{equation}
    \{P\} = \left\{ P \ \middle| \ \begin{gathered} (s,\lambda_\odot)_1 = (s_0,\lambda_{\odot,0}) \\ (s,\lambda_\odot)_{j+1}  \in \mathcal{R}((s,\lambda_\odot)_j) \ \forall (s,\lambda_\odot)_{(j,j+1)} \in P \end{gathered} \right\}.
\end{equation}
% \begin{equation}
% \begin{aligned}
%     J_{\{P\}^*}(s_0,\lambda_{\odot,0}) = \max_{P \in \mathcal{P}(s_0,\lambda_{\odot,0})} \quad & \left\{\begin{matrix}
%         \sum_{(s,\lambda_\odot)_j \in P} Y_j f_{vis}(s_j,\lambda_{\odot,j}) \\
%         -\sum_{(s,\lambda_\odot)_{(j,j+1)} \in P} \Delta V(s_j,\lambda_{\odot,j},s_{j+1},\lambda_{\odot,j+1}) 
%     \end{matrix}\right\} \\
%     \text{s.t.} \quad & (s,\lambda_\odot)_{j+1} \in \mathcal{R}((s,\lambda_\odot)_{j}) \ \forall (s,\lambda_\odot)_{j,j+1} \in P.
%     \end{aligned}
% \end{equation}
\noindent Here, $\mathcal{R}(s,\lambda_{\odot})$ is the set of forward reachable states from $(s,\lambda_{\odot})$, which enforces relevant constraints such that 
\begin{equation}
    \mathcal{R}(s,\lambda_\odot) = \left\{  (s^\prime,\lambda_\odot^\prime) \ \middle| f_{vis}(s^\prime,\lambda_{\odot}^\prime) \neq 0, \
    \lambda_{\odot}^\prime - \lambda_\odot \in [\Delta \lambda_{\odot,min},\Delta \lambda_{\odot,max}]    
    \right\}.
\end{equation}
% Additionally, the set of backward reachable stars can be defined as 
% \begin{equation}
%     \mathcal{R}^{-1}(s,\lambda_\odot) = \left\{  (s^\prime,\lambda_\odot^\prime) \ \middle| f_{vis}(s^\prime,\lambda_{\odot}^\prime) \neq 0, \
%     \lambda_{\odot} - \lambda_\odot^\prime \in [\Delta \lambda_{\odot,min},\Delta \lambda_{\odot,max}]    
%     \right\}.
% \end{equation}
In standard single-objective dynamic programming, the value function would be associated with the optimal objective, such that $V_{P^*} = \max J_{\{P\}}$. Under the multi-objective form, the value function is instead encoded as a Pareto front, i.e., 
\begin{equation}
    V_{\{P\}^*}(s,\lambda_\odot) = \texttt{Pareto}\left\{ J_{\{P\}}(s,\lambda_\odot)\right\}
\end{equation}
\noindent with the Pareto operator defined as $\texttt{Pareto}(X) = \{x \in X \subset \mathbb{R}^2 | \not\exists x^\prime \in X \ \text{s.t.} \ x_1^\prime \ge x_1 \wedge x_2^\prime \ge x_2 \}$. In this form, $V_{\{P\}^*}$ is a discrete list of Pareto optimal objectives and $\{P\}^*$ is a discrete list of the associated Pareto optimal paths. Enumerating over all possible feasible paths is clearly highly combinatorial, however, the principle of dynamic programming can be used to reduce the combinatorial complexity. The associated backwards Bellman update rule in this multi-objective case becomes \cite{MOMDP}
\begin{equation}
    V_{\{P\}^*}(s,\lambda_\odot) = \texttt{Pareto}\left( \bigcup_{(s^\prime,\lambda_\odot^\prime) \in \mathcal{R}((s,\lambda_\odot))} \begin{bmatrix} Y(s) f_{vis}(s,\lambda_\odot) \\ -\Delta V(s^\prime,\lambda_\odot^\prime,s,\lambda_\odot) \end{bmatrix} +V_{\{P\}^*}(s^\prime,\lambda_\odot^\prime) \right)
    \label{eq:BellmanUpdate}
\end{equation}
\noindent where the addition in this expression is element wise with respect to $V_{\{P\}^*}(s^\prime,\lambda_\odot^\prime)$.

\begin{algorithm}[t]
\caption{Observation Scheduling Optimization}\label{alg:SCO}
\vspace{-.3cm}
\begin{flushleft}
\hspace*{\algorithmicindent} \textbf{Input} {Star Catalog $S$, Reachable Set Function $\mathcal{R}(\cdot)$} \\
\hspace*{\algorithmicindent} \textbf{Output} {Value Function $V$, Observation Paths $\{P\}^*$}
\end{flushleft}
\vspace{-.4cm}
\begin{algorithmic}[1]
\Function{\texttt{Scheduling Optimization}}{}
    \State $\{\lambda_\odot\} \gets$ Eq.~\ref{eq:V1} or Eq.~\ref{eq:V2} \Comment{Discretize Time Variable}
    % \State $C \gets$ Eq.~\ref{eq:InitialC} \Comment{Initialize State Set}
    \State $V_{\{P\}^*} \gets$ Eq.~\ref{eq:InitialV} \Comment{Initialize Value Function}
    \For{$\lambda_\odot \in \{ \lambda_{\odot,n}, \cdots,\lambda_{\odot,1} \}$} \Comment{Loop Backwards in Time}
        \For{$s \in \{s \ | \ f_{vis}(s,\lambda_\odot)  \neq 0\}$} \Comment{Loop Each Visible Target}
            % \For{$s \in \mathcal{R}^{-1}(s')$}
            %     \State $V(s') \gets V(s') \bigcup \begin{bmatrix} V_1(s) + R(s') \\ V_2(s) + \Delta V_{s \to s^\prime} \end{bmatrix}$
            % \EndFor
            \State $V_{\{P\}^*}(s,\lambda_\odot) \gets$ Eq.~\ref{eq:BellmanUpdate} \Comment{Optionally, eliminate entries with duplicate observations}
        \EndFor
        % \State $C \gets \mathcal{R}^{-1}(C)$ \Comment{Move to new entries}
    \EndFor
\EndFunction
\end{algorithmic}
\end{algorithm}

Eq.~\ref{eq:BellmanUpdate} defines a backward dynamic programming iteration that can be implemented to compute the set of Pareto-optimal paths. The value function can be initialized as
% Each iteration maintains a current set of states $C = \{(s,\lambda_\odot)_k\}_{k=1}^{n_c}$, which is initialized as
% \begin{equation}
%     C = \{(s,\lambda_\odot) | \lambda_\odot \in [\lambda_{\odot,f} - \Delta \lambda_{\odot,min},\lambda_{\odot,f}],  f_{vis}(s,\lambda_\odot) \neq 0 \},
%     \label{eq:InitialC}
% \end{equation}
% corresponding to the set of states from which no further observations can be made and has associated an associated value function, consisting of the reward and cost of station keeping of the set of states in $C$,
\begin{equation}
    V_{\{P\}^*}(s,\lambda_\odot) = \left\{\begin{bmatrix} Y(s) f_{vis}(s,\lambda_\odot) \\ -\Delta V_{sk}(s,\lambda_{\odot}) f_{vis}(s,\lambda_\odot) \end{bmatrix}\right\} \ \forall s, \lambda_\odot \in (\lambda_{\odot,f}-\Delta \lambda_{\odot,min},\lambda_{\odot,f}].
    \label{eq:InitialV}
\end{equation}
which corresponds to paths that consist of a single observation at the final viewing opportunity, with all other states being initialized as empty. Here, $\lambda_{\odot,0}$ and $\lambda_{\odot,f}$ are the initial and terminal times of the mission duration. An important final implementation detail is the handling of the state variable $\lambda_\odot$, which corresponds to the time of the mission. The method for discretization depends on whether the frozen relative orbit constraint (Eq.~\ref{eq:diyconstraint}) is active or not. In the case that it is, the discretization should take the form 
\begin{equation}
    \{\lambda_\odot\} = \{\lambda_\odot \in [\lambda_{\odot,0},\lambda_{\odot,f}] | \lambda_\odot = \alpha_j + LTAN \pm \pi/2, (\alpha,\delta,Y)_j \in S \}
    \label{eq:V1}
\end{equation}
which naturally results in $(s,\lambda_\odot)$ pairs being such that $\delta i_x = 0$ for the associated relative orbit. An important result of this discretization is the \textit{reduction of the dimension of the state space} to optimize over from two degrees of freedom to one. If Eq.~\ref{eq:diyconstraint} is not active, then the discretization can simply take the uniform form
\begin{equation}
    \{\lambda_\odot\} = \{\lambda_{\odot,1}, \lambda_{\odot,2},\cdots,\lambda_{\odot,n}\} \subset [\lambda_{\odot,0},\lambda_{\odot,f}].
    \label{eq:V2}
\end{equation}
In the ensuing results, this discretization simply consists of an evenly spaced discretization of 1 day intervals. The implementation of these equations in practice is summarized in Alg.~\ref{alg:SCO}. 

One issue with the presented algorithmic approach is the possibility of duplicate observations of the same stellar system. This can only occur in the absence of the $\delta i_x =0$ constraint or when mission durations exceed a duration of 1 year. A simple approach to avoid these scenarios is to simply remove entries that correspond to duplicate observations as part of the Pareto pruning process in Eq.~\ref{eq:BellmanUpdate}. Unfortunately, this has the effect of breaking the Markovian property of the Bellman update, meaning this process introduces optimality losses.

The policy produced provides an optimal sequence of ensuing observation from any initial star and remaining fuel expenditure. This provides much needed flexibility to the observation scheduling scheme, for which any mission may have varying amounts of fuel expenditures or mission interventions from which it may need to proceed. For example, in stellar astronomy, it is common for a star to be reobserved following a detection. In such cases, the scheme naturally provides the optimal path that maximizes scientific reward following this revisit under the remaining $\Delta V$ budget, without requiring a resolve of the optimization problem. This is a novel and desirable aspect of the presented stellar observation scheme.

\section{Results}

Alg.~\ref{alg:SCO} is applied to the star catalog optimization problem of a 6-month LEO interferometer mission for various Sun-synchronous orbits defined in Tab.~\ref{tab:AbsOrbits}. The algorithm provides a set of Pareto-optimal paths of star observations and observation times characterized by $\Delta V$ expenditure and expected exoplanet detections.

\begin{table}[htbp]
\centering
\begin{tabular}{c c c c c c c}
\hline\hline
 & \multicolumn{3}{c}{\textit{Family 1: Terminator}} & \multicolumn{3}{c}{\textit{Family 2: Noon--midnight}} \\
\cline{2-4} \cline{5-7}
 & 1.1 & 1.2 & 1.3 & 2.1 & 2.2 & 2.3 \\
\hline
Altitude [km]     & $500$ & $1000$ & $1500$ & $500$ & $1000$ & $1500$ \\
Inclination       & $97.40^\circ$ & $99.48^\circ$ & $101.96^\circ$ & $97.40^\circ$ & $99.48^\circ$ & $101.96^\circ$ \\
LTAN              & $\pi/2$ & $\pi/2$ & $\pi/2$ & $0$ & $0$ & $0$ \\
Solar Occlusion   & no & no & no & yes & yes & yes \\
\hline\hline
\end{tabular}
\caption{Absolute orbit test scenarios for catalog optimization. For Family 2, observations are performed solely during times at which the Sun is occluded by the Earth (see Eq.~\ref{eq:SolarOccultation}). Each case is run with the addition of the $\delta i_x = 0$ constraint (Case a.) and without the addition of the $\delta i_x = 0$ constraint (Case b.).}
\label{tab:AbsOrbits}
\end{table}

\begin{figure}[!htbp]
    \centering

    % Left column
    \begin{minipage}[c]{0.48\textwidth}
        \centering
        \begin{subfigure}{\linewidth}
            \centering
            \includegraphics[width=2.9in]{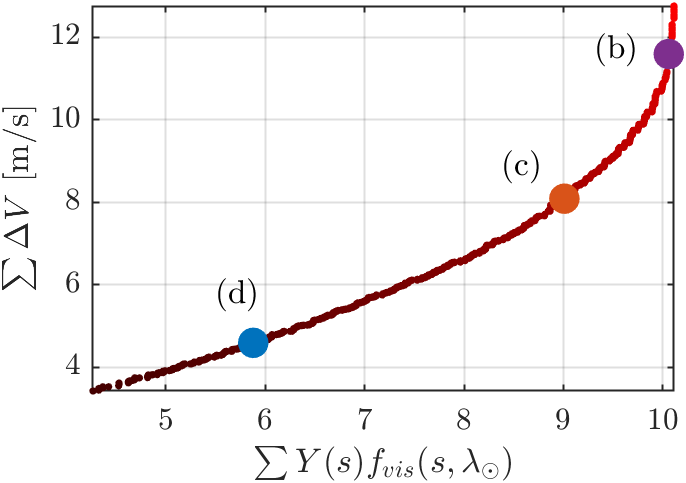}\\[.2in]
            \includegraphics[width=2.9in]{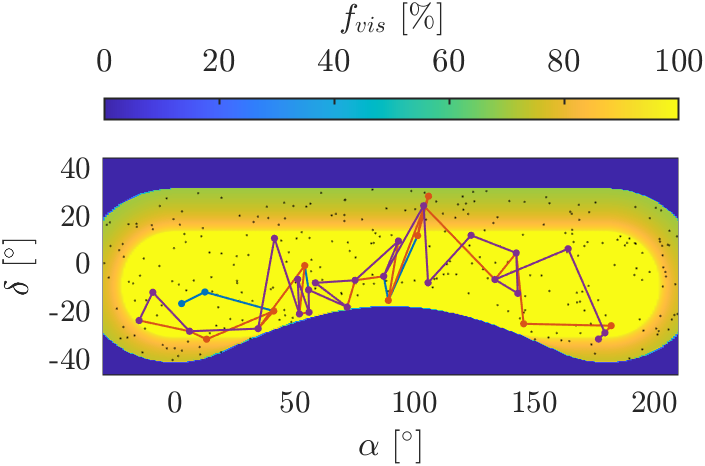}\\[.2in]
            \includegraphics[width=2.9in]{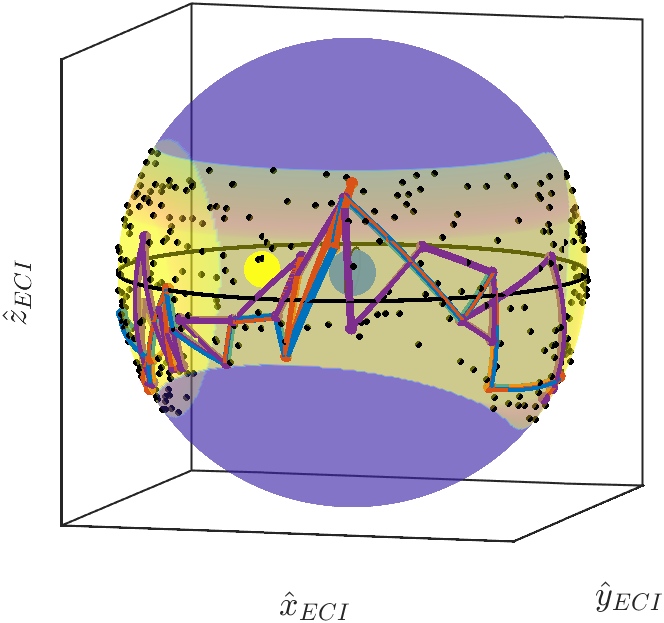}
            \caption{Pareto front and paths}
            \label{fig:Mainpar}
        \end{subfigure}
    \end{minipage}
    \hfill
    % Right column
    \begin{minipage}[c]{0.48\textwidth}
        \centering

        \begin{subfigure}{\linewidth}
            \centering
            \includegraphics[width=1.4in]{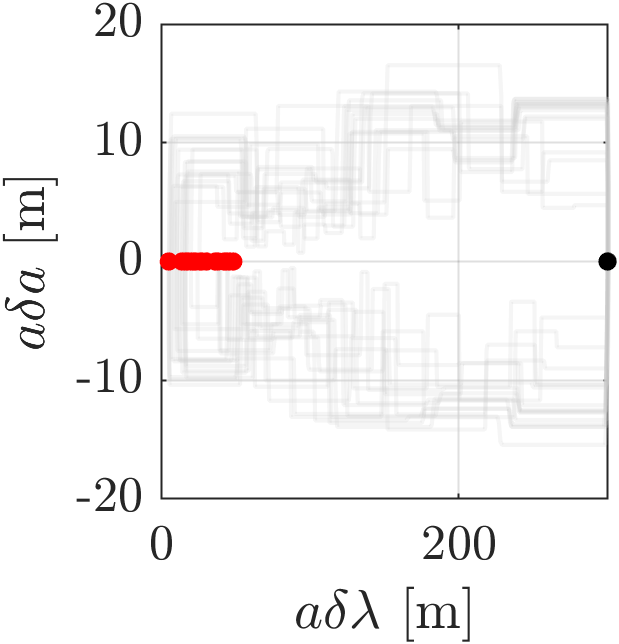}
            \includegraphics[width=1.4in]{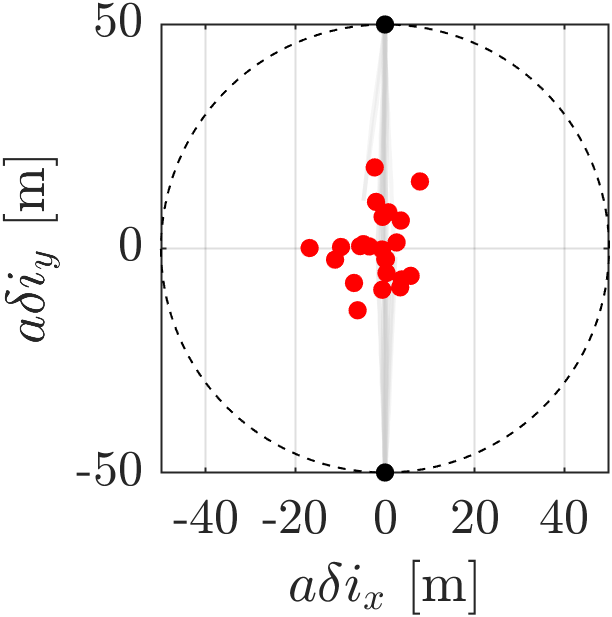}
            \includegraphics[width=2.8in]{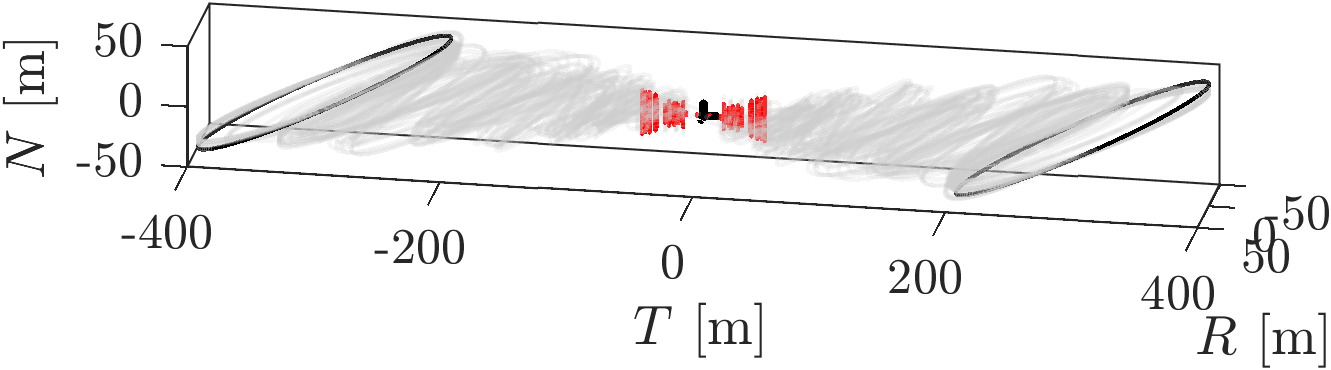}
            \caption{High-Fuel, High-Reward}
            \label{fig:traj1}
        \end{subfigure}

        \vspace{0.5em}

        \begin{subfigure}{\linewidth}
            \centering
            \includegraphics[width=1.4in]{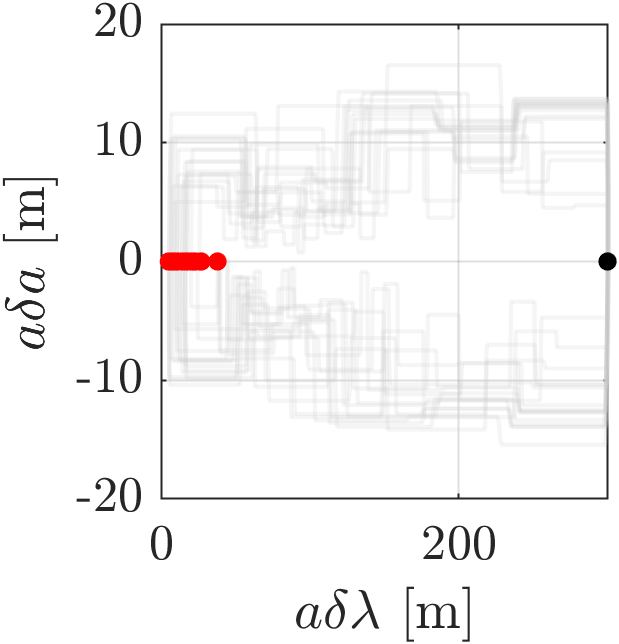}
            \includegraphics[width=1.4in]{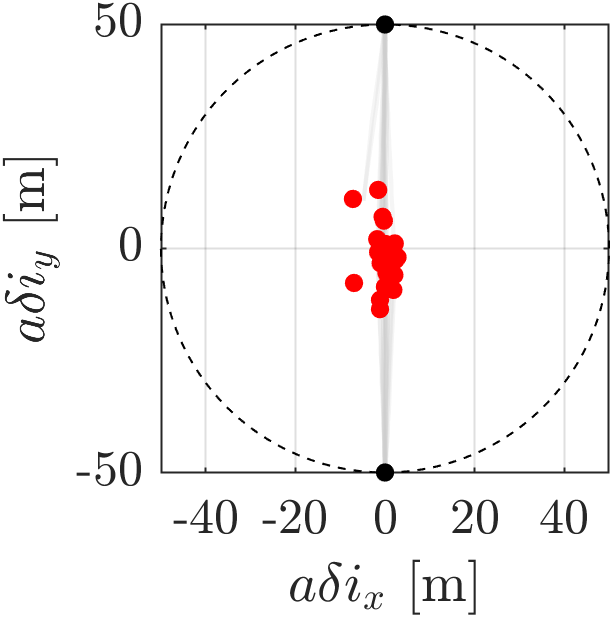}
            \includegraphics[width=2.8in]{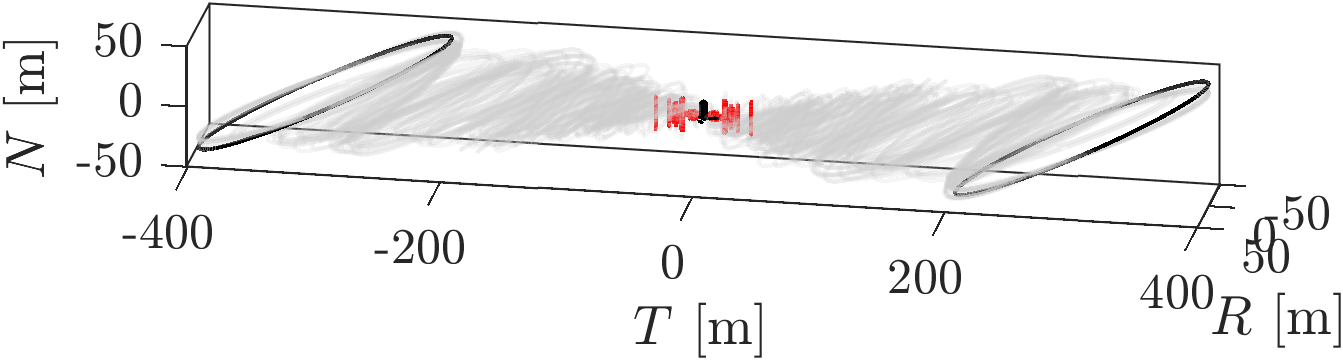}
            \caption{Medium-Fuel, Medium-Reward}
            \label{fig:traj2}
        \end{subfigure}

        \vspace{0.5em}

        \begin{subfigure}{\linewidth}
            \centering
            \includegraphics[width=1.4in]{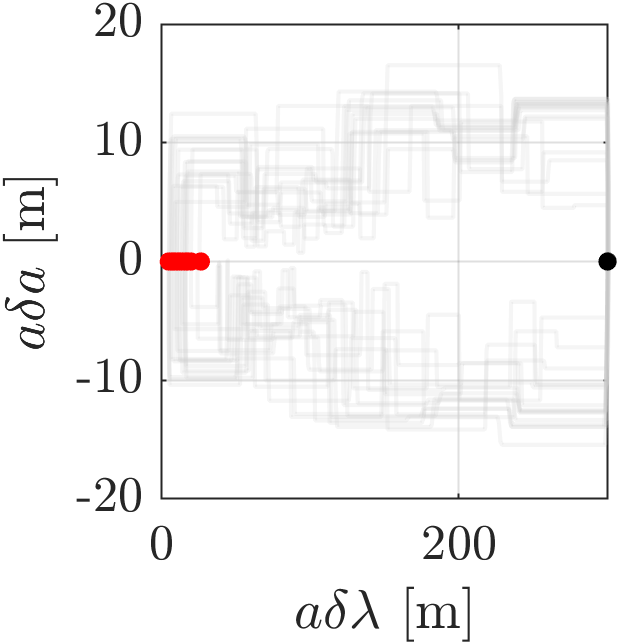}
            \includegraphics[width=1.4in]{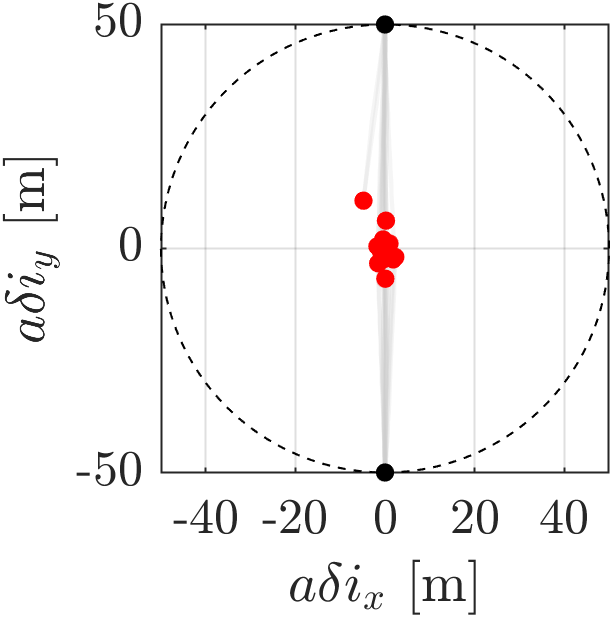}
            \includegraphics[width=2.8in]{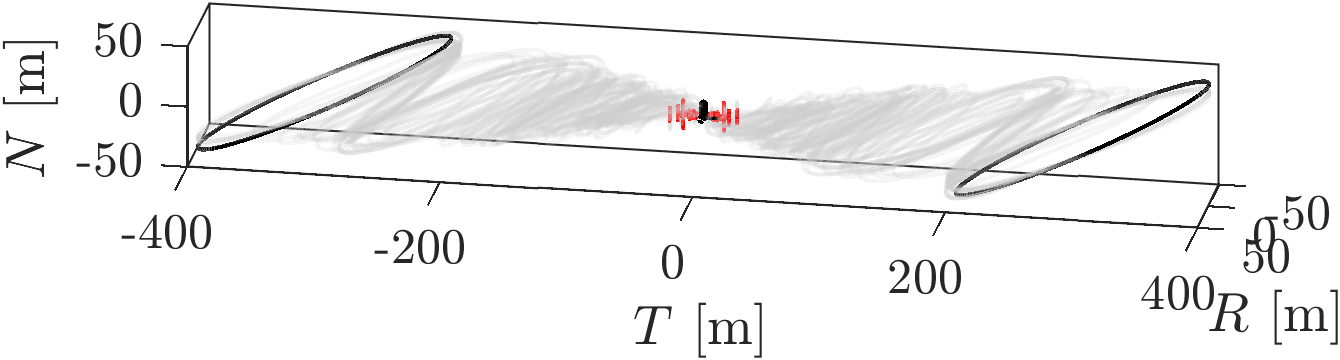}
            \caption{Low-Fuel, Low-Reward}
            \label{fig:traj3}
        \end{subfigure}
    \end{minipage}

    \caption{Pareto-optimal observation schedule for a 6-month LEO interferometry mission (Case 1.1b of Tab.~\ref{tab:AbsOrbits}). Fig.~\ref{fig:Mainpar} shows the Pareto front from the mission start time and three selected points on the Pareto front whose paths are plotted in the equatorial sky map and celestial sphere. The associated trajectories, consisting of science orbits (red), standby orbits (black), and transfers between (gray), in relative orbital element $\delta a-\delta \lambda$ and $\delta i_x -\delta i_y$ coordinates and RTN frame positions are shown in Figs.~\ref{fig:traj1}, \ref{fig:traj2}, and \ref{fig:traj3} for each of the three paths. The Standby orbit circle is plotted in the $\delta \bm{i}$ plane.}
    \label{fig:MainTraj}
\end{figure}

Fig.~\ref{fig:MainTraj} displays the Pareto front, path plans, and spacecraft trajectories for a 6-month LEO interferometry mission stationed in a Sun-synchronous terminator orbit associated with Case 1.1b. $\Delta V$ expenditure ranges from $\sim 3.5-13$ m/s and yield ranges from $\sim 4-10$ expected exoplanet detections. The paths move from left to right on the right-ascension ($\alpha$) and declination ($\delta$) plane sky map in Fig.~\ref{fig:Mainpar} as a result of the drift of the Sun-synchronous orbit over the 6-month duration. This conveniently keeps the Sun exclusion zone consistent with other visibility constraints like maximum baseline oscillation, therefore maximizing stellar target visibility over the mission duration. This sun exclusion also has the effect of forbidding observations of lower declination stellar targets around $\alpha \sim 50^\circ -130^\circ$ during winter months as a result of the suns obliquity with respect to the equatorial plane. 

Analyzing Figs.~\ref{fig:traj1}, \ref{fig:traj2}, and \ref{fig:traj3} further, the trajectories show a trend in which lowering $\Delta V$ expenditure along the Pareto front is associated with science orbits with smaller $\delta \bm{i}$ and $\delta \lambda$, which is attributable to the fact that these science orbits incur lower station keeping costs (Eq.~\ref{eq:SK}). Moreover, the low-fuel, low-reward trajectory consists of observing 2 G-class, 1 K-class, and 13 M-class stellar systems, the medium-fuel, medium-reward trajectory consists of observing 3 F-class, 2 G-class, and 18 M-class stellar systems, and the high-fuel, high-reward trajectory consists of observing 10 F-class, 1 G-class, 11 M-class, and 1 K-class stellar system. 

The scheme is also used to compare mission concepts arising from a variety of absolute orbit conditions. This includes six different absolute orbits of interest with varying inclinations, altitudes, and LTANs, as summarized in Tab.~\ref{tab:AbsOrbits}. 

% \begin{table}[htbp]
% \centering
% \begin{tabular}{c c c c c}
% \hline\hline
% Case & Altitude [km] & Inclination & LTAN & Solar Occlusion \\
% \hline
% \multicolumn{5}{c}{\textit{Family 1: Sun-synchronous, terminator orbit}} \\
% 1.1 & $500$  & $97.40^\circ$ & $\pi/2$ & no \\
% 1.2 & $1000$ & $99.48^\circ$ & $\pi/2$ & no \\
% 1.3 & $1500$ & $101.96^\circ$ & $\pi/2$ & no \\
% \hline
% \multicolumn{5}{c}{\textit{Family 2: Sun-synchronous, noon--midnight orbit}} \\
% 2.1 & $500$  & $97.40^\circ$ & $0$ & yes \\
% 2.2 & $1000$ & $99.48^\circ$ & $0$ & yes \\
% 2.3 & $1500$ & $101.96^\circ$ & $0$ & yes \\
% \hline
% \multicolumn{5}{c}{\textit{Family 3: Equatorial orbit}} \\
% 3.1 & $500$  & $0^\circ$ & --- & yes \\
% 3.2 & $1000$ & $0^\circ$ & --- & yes \\
% 3.3 & $1500$ & $0^\circ$ & --- & yes \\
% \hline\hline
% \end{tabular}
% \caption{Absolute orbit test scenarios for catalog optimization. For Families 2 and 3, observations are performed solely during times at which the Sun is occluded by the Earth (see Eq.~\ref{eq:SolarOccultation}). Each case is run with the addition of the $\delta i_x = 0$ constraint (Case a.) and without the addition of the $\delta i_x = 0$ constraint (Case b.).}
% \label{tab:AbsOrbits}
% \end{table}

\begin{figure}[!htbp]
    \centering

    %==================== Row 1 ====================

    \begin{subfigure}{2.9in}
        \centering
        \includegraphics[width=1.1in]{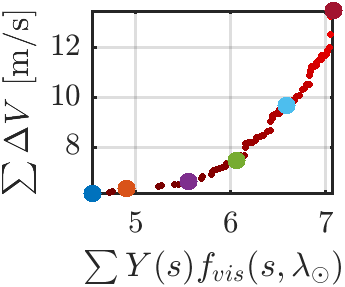}
        \hfill
        \includegraphics[width=1.7in]{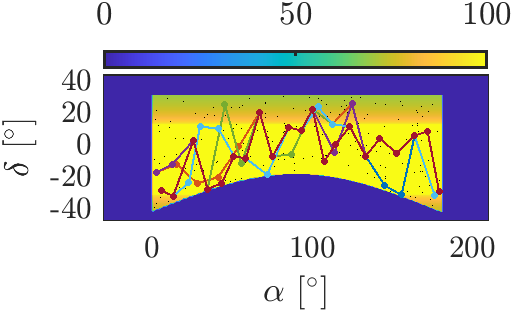}
        % \caption{Case 1.1a.}
        \label{fig:1.1a}
    \end{subfigure}
    \hfill
    \begin{subfigure}{2.9in}
        \centering
        \includegraphics[width=1.1in]{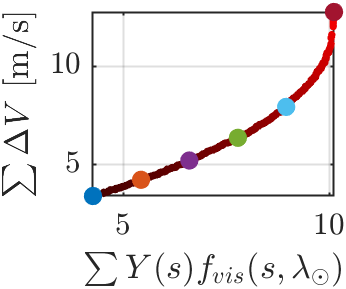}
        \hfill
        \includegraphics[width=1.7in]{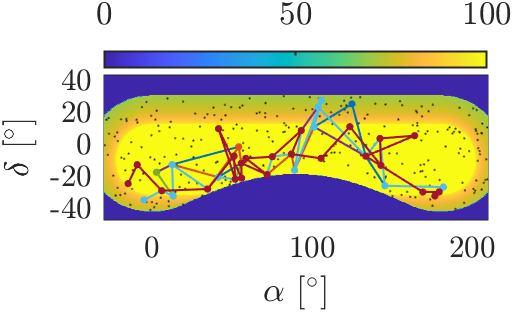}
        % \caption{Case 1.1b}
        \label{fig:1.1b}
    \end{subfigure}

    \begin{subfigure}{2.9in}
        \centering
        \includegraphics[width=1.1in]{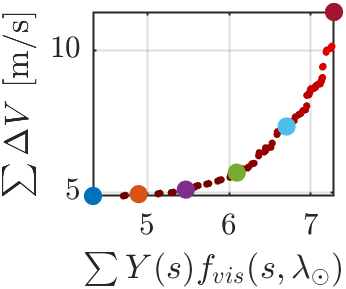}
        \hfill
        \includegraphics[width=1.7in]{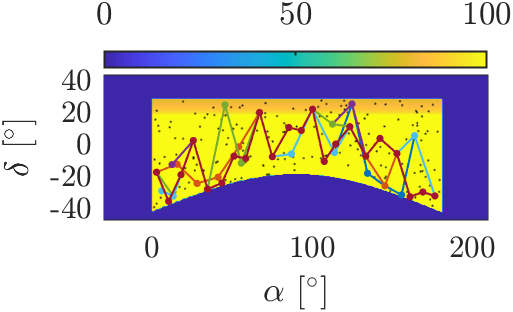}
        % \caption{Case 1.2a.}
        \label{fig:1.2a}
    \end{subfigure}
    \hfill
    \begin{subfigure}{2.9in}
        \centering
        \includegraphics[width=1.1in]{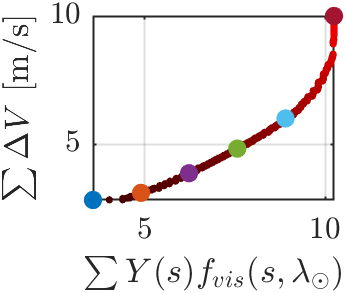}
        \hfill
        \includegraphics[width=1.7in]{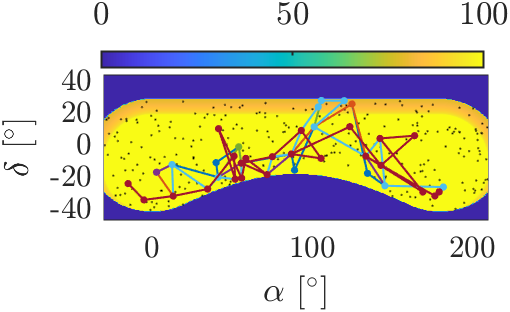}
        % \caption{Case 1.2b}
        \label{fig:1.2b}
    \end{subfigure}

    \begin{subfigure}{2.9in}
        \centering
        \includegraphics[width=1.1in]{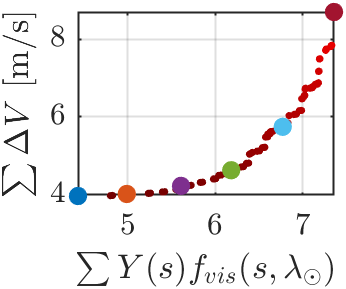}
        \hfill
        \includegraphics[width=1.7in]{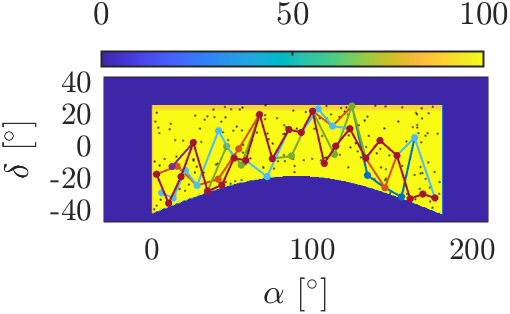}
        \caption{Cases 1.1a., 1.2a., and 1.3a.}
        \label{fig:1.3a}
    \end{subfigure}
    \hfill
    \begin{subfigure}{2.9in}
        \centering
        \includegraphics[width=1.1in]{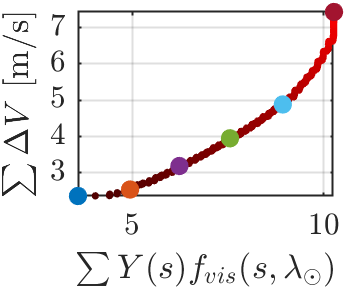}
        \hfill
        \includegraphics[width=1.7in]{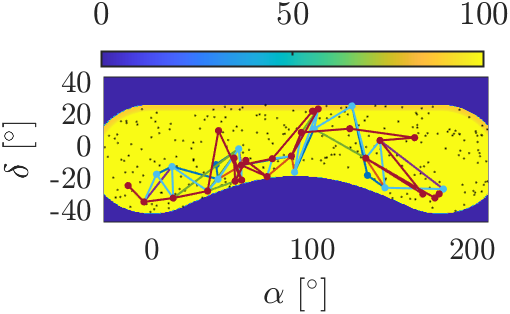}
        \caption{Cases 1.1b., 1.2b., and 1.3b.}
        \label{fig:1.3b}
    \end{subfigure}

    \caption{Pareto fronts and paths of Family 1: Sun-synchronous, terminator orbit. Fig.~\ref{fig:1.3a} considers a $\delta i_x =0$ constraint whereas Fig.~\ref{fig:1.3b} does not. Top, middle, and bottom figures correspond to $500$, $1000$, and $1500$ km altitude orbits, respectively. Color bars denote visibility percentage, $f_{vis}$ [\%].}
    \label{fig:F1}
\end{figure}

\begin{figure}[!htbp]
    \centering

    \begin{subfigure}{2.9in}
        \centering
        \includegraphics[width=1.1in]{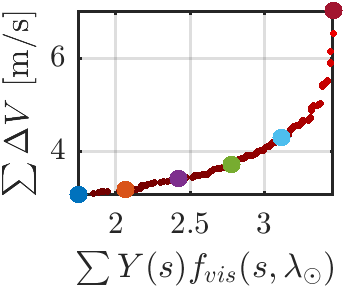}
        \hfill
        \includegraphics[width=1.7in]{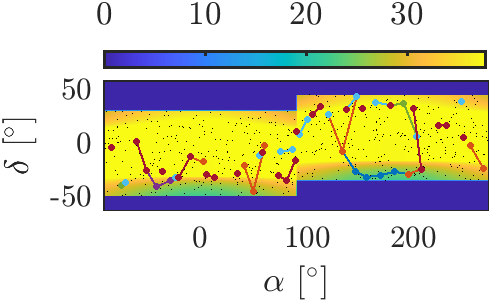}
        % \caption{Case 2.1a.}
        \label{fig:2.1a}
    \end{subfigure}
    \hfill
    \begin{subfigure}{2.9in}
        \centering
        \includegraphics[width=1.1in]{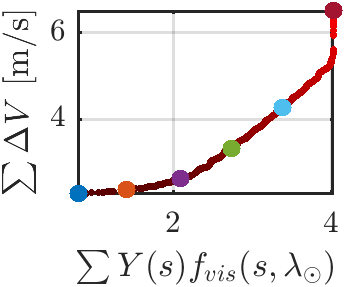}
        \hfill
        \includegraphics[width=1.7in]{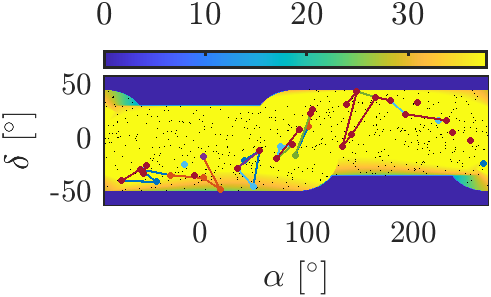}
        % \caption{Case 2.1b}
        \label{fig:2.1b}
    \end{subfigure}

    \begin{subfigure}{2.9in}
        \centering
        \includegraphics[width=1.1in]{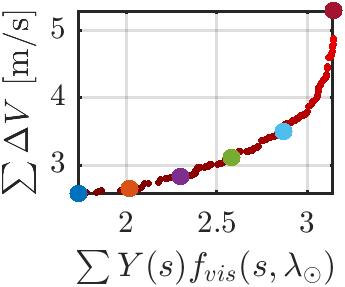}
        \hfill
        \includegraphics[width=1.7in]{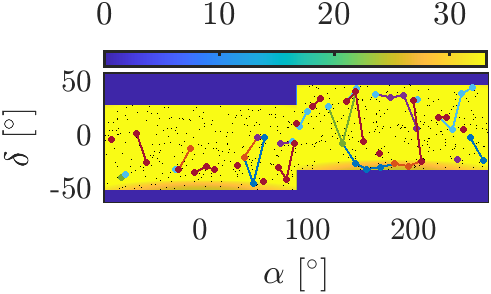}
        % \caption{Case 2.2a.}
        \label{fig:2.2a}
    \end{subfigure}
    \hfill
    \begin{subfigure}{2.9in}
        \centering
        \includegraphics[width=1.1in]{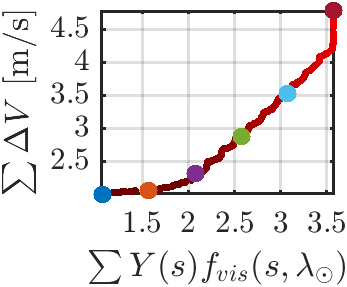}
        \hfill
        \includegraphics[width=1.7in]{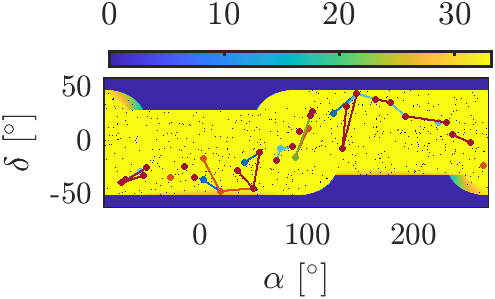}
        % \caption{Case 2.2b}
        \label{fig:2.2b}
    \end{subfigure}

    \begin{subfigure}{2.9in}
        \centering
        \includegraphics[width=1.1in]{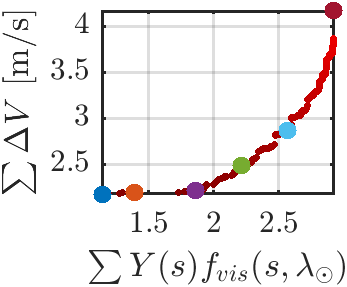}
        \hfill
        \includegraphics[width=1.7in]{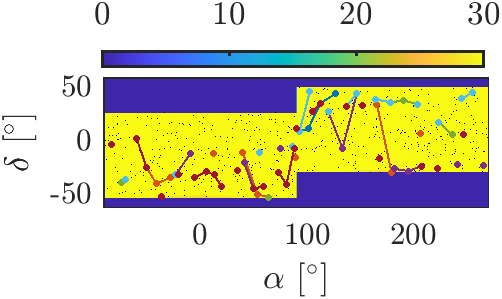}
        \caption{Cases 2.1a., 2.2a., and 2.3a.}
        \label{fig:2.3a}
    \end{subfigure}
    \hfill
    \begin{subfigure}{2.9in}
        \centering
        \includegraphics[width=1.1in]{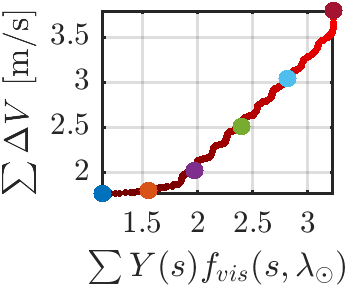}
        \hfill
        \includegraphics[width=1.7in]{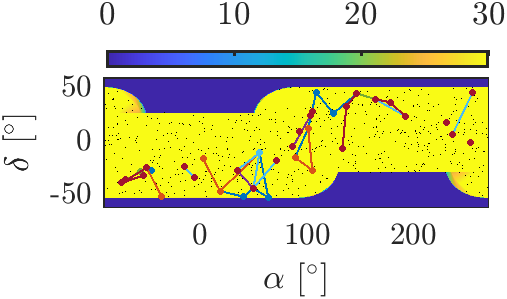}
        \caption{Cases 2.1b., 2.2b., and 2.3b.}
        \label{fig:2.3b}
    \end{subfigure}

        \caption{Pareto fronts and paths of Family 2: Sun-synchronous, noon-midnight orbit. Fig.~\ref{fig:2.3a} considers a $\delta i_x =0$ constraint whereas Fig.~\ref{fig:2.3b} does not. Top, middle, and bottom figures correspond to $500$, $1000$, and $1500$ km altitude orbits, respectively. Breaks in the path lines of the sky map figures denote switches to the opposite side to the observation plane, i.e., when $\alpha_{j+1} \approx \alpha_j \pm 180^\circ$.}
    \label{fig:F2}
\end{figure}

\begin{figure}[!htb]
    \centering
    \begin{minipage}[c]{0.63\textwidth}
    \centering
    \includegraphics[width=3in]{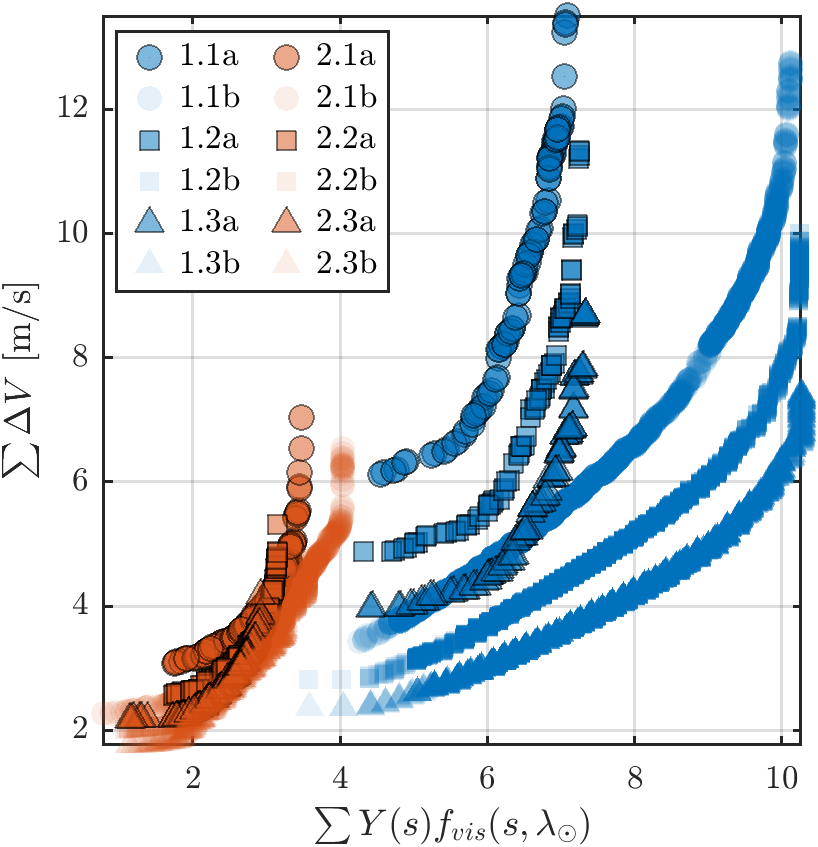}
\end{minipage}
\hspace{-.2in}
\begin{minipage}[c]{0.32\textwidth}
    \centering
    \includegraphics[width=1.5in]{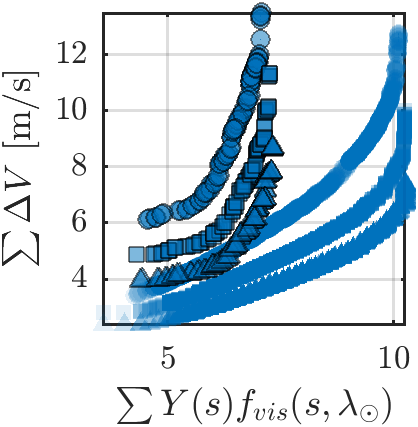}

    \vspace{0.15in}

    \includegraphics[width=1.5in]{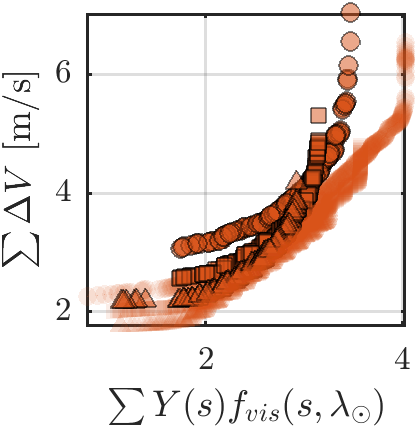}
\end{minipage}
    \caption{Comparison of Pareto fronts for each case in Tab.~\ref{tab:AbsOrbits}. Blue markers are the orbits of Family 1 (terminator orbits), whereas orange markers are the orbits of Family 2 (noon-midnight orbits). The shape of the markers distinguish orbital altitude and the edge shading distinguish the presence of the $\delta i_x = 0$ constraint (Case a. vs. b.).}
    \label{fig:CatOptParetoComparison}
    \vspace{-.2in}
\end{figure}

The resulting Pareto fronts and paths for each case are visualized in Figs.~\ref{fig:F1} and \ref{fig:F2}. For each Pareto front, six paths are selected and plotted in the stellar target right ascension and declination plane. For each of these planes, the paths are underlaid by the available viewing region associated with the mission parameters defined in Tab.~\ref{tab:AbsOrbits}. For Family 1 (Fig.~\ref{fig:F1}), visibility of the celestial sphere improves as the orbital altitude increases as a result of lower Earth occlusion percentages. However, for Family 2 (Fig.~\ref{fig:F2}), visibility of the celestial sphere degrades as the orbital altitude increases as a result of lower solar occlusion percentages. The absence of the $\delta i_x = 0$ constraint (case b.) also produces greater accessibility of the celestial sphere at the start and end of the mission. The absolute orbits of Family 1 (Fig.~\ref{fig:F1}) restrict observations to $\sim\alpha \in [0^\circ,180^\circ]$ as a result of the Sun exclusion constraint, which forbids observations of the opposite side of the celestial sphere for the 6-month mission duration. In contrast, the absolute orbits of Family 2 (Fig.~\ref{fig:F2}) feature observations across the full range of $\alpha$. This is a result of the solar occlusion constraint permitting viewing of opposite sides of the celestial sphere at any given time.  

In addition, the Pareto fronts from each case are compared in Fig.~\ref{fig:CatOptParetoComparison}. For Family 1, displayed in blue, increasing the altitude of the absolute orbit leads to uniformly lower $\Delta V$ expenditure for equivalent scientific reward. This is a result of the lower $\Delta V$ needed for transfer and station keeping at these higher altitudes. Moreover, the absence of the $\delta i_x = 0$ constraints (Cases a. vs. b.) leads to improved scientific potential and reduced $\Delta V$ expenditure as a result of the fact that the reachable space in the absence $\delta i_x = 0$ constraint is far less restricted. In contrast, Family 2, displayed in orange, features significantly reduced scientific performance compared to Family 1, which is attributable to the lower visibility fraction $f_{vis}$ for this family of orbits. Stellar targets have on average half as much visibility for orbits of Family 2 compared to Family 1 as a result of observations taking place only when the Sun is occluded by the Earth. In addition, this visibility fraction is decreased with increasing altitude, leading to lower scientific potential with higher altitude orbits.

\section{Conclusion}

This paper proposed a multi-objective dynamic programming scheme for determining stellar observation scheduling for future linear nulling interferometer concepts. The consideration of realistic mission constraints and performance drivers, such as fuel expenditure, scientific yield, and observability constraints, lends realism and real-world applicability of the scheme to future missions and is, in addition, a useful tool for comparing how various design parameters affect scientific potential of a mission. The global policy that is produced lends flexibility to changes in mission profiles over time, such as the occurrence of revisits, and is a unique approach for scheduling optimization in astronomy literature. Moreover, the scheme is a useful tool for developing a flexible concept of operations for the mission profile. The presented results suggest that LEO-based nulling interferometers can achieve a strong scientific yield of up to 10 expected protoplanet detections with modest mission profiles of meters to 10s of meters per second of $\Delta V$ over a 6-month mission duration. Future iterations may seek to improve the metric of scientific reward and extend the scheme to eccentric and Sun-Earth $L_2$ orbital regimes.

\section{Acknowledgements}

This research was developed in association with the STARI mission under the NASA-APRA program number 80NSSC25K7101, which is coordinated by the University of Michigan with Prof. John Monnier as Principal Investigator. The authors thank Yuji Takubo and Shane Lowe for insightful discussions and advice.

\newpage

\section*{Appendix: Science Yield Computations} \label{app:Science}

\begin{table*}[t]
\centering
\renewcommand{\arraystretch}{0.95}
\setlength{\tabcolsep}{3pt}
\begin{minipage}{0.48\textwidth}
\centering
\begin{tabular}{@{}lll@{}}
\hline\hline
Parameter & Symbol & Value \\
\hline
\multicolumn{3}{l}{\textit{Wavelength band}} \\
Central wavelength           & $\lambda_0$              & $10\ \mu\mathrm{m}$ \\
Lower band edge              & $\lambda_{\min}$       & $4\ \mu\mathrm{m}$ \\
Upper band edge              & $\lambda_{\max}$       & $18.5\ \mu\mathrm{m}$ \\
\hline
\multicolumn{3}{l}{\textit{Background sources}$^a$} \\
Local zodi brightness        & $I_{\mathrm{zodi}}$    & $1\ \mathrm{\frac{MJy}{sr}}$ \\
Local zodi temperature & $T_{{zodi}}$ & $265$ K \\
Exozodi brightness           & $I_{\mathrm{exo}}$     & $3\ \mathrm{\frac{MJy}{sr}}$ \\
Exozodi temperature & $T_{{exo}}$ & $265$ K \\
\hline
\multicolumn{3}{l}{\textit{Array configuration}} \\
Total Area & $A_{tot}$ & 1.57 $\mathrm{m}^2$ \\
Minimum baseline             & $B_{\min}$             & $5\ \mathrm{m}$ \\
Maximum baseline             & $B_{\max}$             & $200\ \mathrm{m}$ \\
Total throughput             & $\eta$                 & $0.05$ \\
\hline\hline
\end{tabular}
\end{minipage}
\hfill
\begin{minipage}{0.48\textwidth}
\centering
\begin{tabular}{@{}lll@{}}
\hline\hline
Parameter & Symbol & Value \\
\hline
\multicolumn{3}{l}{\textit{Habitable zone\cite{kopparapuHabitableZonesMainSequence2013}}} \\
Inner HZ flux limit          & $S_{\mathrm{in}}$      & $1.1\ S_\oplus$ \\
Outer HZ flux limit          & $S_{\mathrm{out}}$     & $0.53\ S_\oplus$ \\
\hline
\multicolumn{3}{l}{\textit{Target planet \& detection}} \\
Planet radius                & $R_p$                  & $1.0\ R_\oplus$ \\
Planet temperature           & $T_p$                  & $600\ \mathrm{K}$ \\
SNR threshold \cite{Quanz_2022}      & $SNR$         & $7$ \\
Max.\ integration time       & $t_{{eff,max}}$ & $1\ \mathrm{day}$ \\
Monte Carlo samples          & $N_{{mc}}$      & $1000$ \\
\hline
\multicolumn{3}{l}{\textit{HZ occurrence rates}$^b$} \\
F-type host                  & $\eta_{{F}}$    & $0.30$ \\
G-type host                  & $\eta_{{G}}$    & $0.50$ \\
K-type host                  & $\eta_{{K}}$    & $0.45$ \\
M-type host                  & $\eta_{{M}}$    & $0.20$ \\
\hline\hline
\end{tabular}
\end{minipage}
\vspace{4pt}

\begin{minipage}{\textwidth}
\footnotesize
\begin{tabular}{@{}p{0.02\textwidth}p{0.9\textwidth}@{}}
{[a]} & Calibrated against Ref.~\citenum{dannertLargeInterferometerExoplanets2022} (Fig.~5). \\\relax
{[b]} & Assigned by host spectral type: G from Ref.~\citenum{Bryson_2021}, M from Ref.~\citenum{dressingOccurrencePotentiallyHabitable2015}.
\end{tabular}
\end{minipage}
\caption{Model parameters for mid-infrared nulling interferometer yield.}
\label{tab:ScienceParameters}
\end{table*}

Stellar target yield is computed as a function of stellar target data provided by the Habitable Worlds Observatory stellar target catalog \cite{Tuchow_2025}. First, several system types are removed, including known binary systems, systems with a parallax quality error greater than 20\%, systems with ambiguous distances, contaminated systems, and systems that are not dwarfs of Morgan-Keenan class F, G, K or M. Each stellar system has a given stellar distance in parsecs, $D_*$, stellar radius $R_*$, stellar temperature in Kelvin, $T_{*}$, and stellar luminosity, $L_*$, in Watts. From this, the inner and outer habitable zone radii in astronomical units (AU) are $a_{in} = \sqrt{(L_*/L_\odot)/S_{in}}$ and $a_{out} = \sqrt{(L_*/L_\odot)/S_{out}}$ respectively. The angular separation for any exoplanet with semi-major axis $a\in [a_{in},a_{out}]$ is $\theta_{HZ} = a/D_*$, with units of radians. The baseline $\theta_{HZ}$ is selected as the average of $a_{in}$ and $a_{out}$. Similarly, the stellar angular diameter is $\theta_* = 2R_*/D_*$.

% The optimal baseline of the linear array is chosen such that $B_0 = 0.59 \frac{\lambda_0}{\theta_{HZ}}$ \cite{hansenPreliminaryExplorationEffects2026}, approximately placing the HZ on the central nulling peak. This value is also clipped such that $B_0 \in [B_{min},B_{max}]$. This corresponds to an inner working angle $IWA = 0.5 \frac{\lambda_0}{B_0}$.

To simulate sample rates via Monte Carlo, the exoplanet distance from the stellar host can be sampled via $r_j = a_j \sqrt{\cos^2 \phi + \sin^2 \phi \cos^2 \iota}$, where $a_j \sim \mathcal{U}[a_{in},a_{out}]$, $\phi \sim \mathcal{U}[-\pi,\pi]$, and $\cos \iota \sim \mathcal{U}[-1,1]$. The associated angular separations are $\theta_j = r_j/D_*$. 

Spectral irradiance is used to compute signal strengths of various objects, including the host star, exoplanet, and zodiacal dust. Each make use of the Planck photon radiance formula $B_{ph}(\lambda,T) = \frac{2c}{\lambda^4}/ \left(\exp\frac{hc}{\lambda k_B T} - 1\right)$, where $c$ is the speed of light, $h$ is the Planck constant, $k_B$ is the Boltzmann constant, and $\lambda$ and $T$ are the light wavelength and temperature of the object respectively.  For each signal source, the spectral irradiances are 
\begin{equation}
    \begin{gathered}
        E_p(\lambda) =\pi \left( \frac{R_p}{D_*} \right)^2 B_{{ph}}(\lambda, T_p), \
        E_*(\lambda) =\pi \left( \frac{R_*}{D_*} \right)^2 B_{{ph}}(\lambda, T_*), \\ 
        E_{{zodi}}(\lambda) = \frac{I_{{zodi}}}{B_{{ph}}(\lambda_0,T_{{zodi}})\lambda_0} B_{{ph}}(\lambda,T_{{zodi}}), \  
        E_{{exo}}(\lambda) = \frac{2I_{{exo}}}{B_{{ph}}(\lambda_0,T_{{exo}})\lambda_0} B_{{ph}}(\lambda,T_{{exo}}).
    \end{gathered}
\end{equation}
For a single-Bracewell ($\theta^2$, unchopped) architecture, the signal strengths in photons per second are \cite{absilAstrophysicalStudiesExtrasolar,dannertLargeInterferometerExoplanets2022}
\begin{equation}
    \begin{gathered}
        R_{{sig}} = \eta A_{{tot}} \int_{\lambda_{{min}}}^{\lambda_{{max}}} E_p(\lambda) \sin^2(\pi \theta_j B_0/\lambda) d \lambda, \\ R_{{leak}} = \eta A_{{tot}} \int_{\lambda_{{min}}}^{\lambda_{{max}}} E_*(\lambda) \min\left(\frac{\pi^2}{16}(\theta_* B_0/\lambda)^2,1\right) d \lambda , \\
        R_{{zodi}} = \eta \int_{\lambda_{{min}}}^{\lambda_{{max}}} E_{{zodi}}(\lambda) \lambda^2 d \lambda, \  R_{{exo}} = \eta \int_{\lambda_{{min}}}^{\lambda_{{max}}} E_{{exo}}(\lambda) \lambda^2 d \lambda. 
    \end{gathered}
\end{equation}
These integrals can be evaluated using numerical quadrature. From this, the integration time $t_{{eff},j}$ to exceed the SNR threshold can be computed using Eq.~\ref{eq:SNR}. Finally, the completeness of the system is computed with Eq.~\ref{eq:Completeness}. Relevant numerical values can be found in Tab.~\ref{tab:ScienceParameters}

\section*{Appendix: Relative Orbit Transfers with Standby Relaxation} \label{app:Transfer}

The presented studies consider standby orbits which are restricted to the $\delta i_y$ axis. However, this can be generalized to the full $\delta \bm{i}$ plane if desired as follows. Using the previously presented analytical bound,
\begin{equation}
    \Delta V = an (\| \Delta \delta \bm{e} \|/2 + \| \Delta \delta \bm{i} \| )
\end{equation} 
the $\Delta V$ cost of transitioning between two observations whose associated locations on the relative inclination plane are $\delta \bm{i}_0$ and $\delta \bm{i}_f$ is given by
\begin{equation}
\begin{aligned}
    \Delta V_{0 \rightarrow f} &= an \left( \min_{a \min\{ \| \delta \bm{i}_{sb}\|,\| \delta \bm{e}_{sb} \| \} \ge r_{min}, \delta \bm{i}_{sb} || \delta \bm{e}_{sb}} \|\delta \bm{i}_0-\delta \bm{i}_{sb} \| + \| \delta \bm{i}_f - \delta \bm{i}_{sb} \| + \| \delta \bm{e}_{sb} \| \right) \\
    &= n \left( r_{min} + a\min_{a \| \delta \bm{i}_{sb}\|\ge r_{min}} \|\delta \bm{i}_0-\delta \bm{i}_{sb} \| + \| \delta \bm{i}_f - \delta \bm{i}_{sb} \|\right)
    \end{aligned}
\end{equation}

The minimizer to this sequence is in fact an instance of Alhazen's problem, and consists of minimizing the cumulative distance between two points such that the path is coincident with the standby radius at some point. The point on the standby circle has a solution form based on the root of a quartic polynomial \cite{Adeceptivelyeasyproblem}, with
\begin{equation}
    \overline{(\delta i_1 \delta i_2) } \delta i_{sb}^4-(\bar{\delta i_1}+\bar{\delta i_2}) \delta i_{sb}^3+(\delta i_1+\delta i_2) \delta i_{sb}-\delta i_1 \delta i_2=0
\end{equation}
\noindent where $\delta i_1 = (\delta i_{x,0} + \delta i_{y,0} \sqrt{-1})/r_{min} \in \mathbb{C}$ and $\delta i_2 = (\delta i_{x,f} + \delta i_{y,f} \sqrt{-1})/r_{min} \in \mathbb{C}$ are the normalized complex versions of the relative inclination vectors, and $\delta i_{sb} \in \mathbb{C}$ is the normalized complex version of the corresponding minimizing standby configuration. Though this polynomial gives multiple roots, one of them is guaranteed to be the minimizing argument on the unit circle, which can be used to compute the associated fuel expenditure of the transfer. 

\bibliographystyle{AAS_publication}   % Number the references.
\bibliography{references}   % Use references.bib to resolve the labels.

\end{document}